\documentclass[preprint,showpacs,showkewords,preprintnumber,amsmath,amssymb]{revtex4}
 \usepackage{tipa}
 \usepackage{amsmath}
 \usepackage{txfonts}
 \usepackage{amssymb}
 \usepackage{graphicx,epsfig}
 \usepackage{bm}
 \begin{document}

 \title{ Lepton mixing patterns from combinations of elementary correlations}
 \author{Shu-jun Rong}\email{rongshj@snut.edu.cn}

 \affiliation{Department of Physics, Shaanxi University of Technology, Hanzhong, Shaanxi 723000, China}

 \begin{abstract}
Recent data of reactor neutrino experiments set more stringent constraints on leptonic mixing  patterns. We examine all possible patterns on the basis of combinations of elementary correlations of elements of leptonic mixing matrix.  We obtain 62 viable mixing patters at 3$\sigma$ level of mixing parameters. Most of these patterns can be paired via the ¦Ì- interchange which changes the octant of $\theta_{23}$ and the sign of $\cos{\delta}$.  All viable patterns can be classified into two groups: the perturbative patterns and nonperturbative patterns. The former can be obtained from  perturbing TBM. The latter cannot be obtained from  perturbing any mixing pattern whose $\theta_{13}$ is zero. Different predictions of Dirac CP phase $\delta$ of these two types of mixing patterns are discussed. Evolutions of mass matrices of neutrinos with small mixing parameters are discussed via special mixing patterns on the basis of flavor groups.  In general cases, a small variation of $\sin\theta_{13}$
may bring about large modifications to alignment of vacuum expectation values in a mixing model. Therefore, small but nonzero $\sin\theta_{13}$ brings a severer challenge to leptonic mixing models on the basis of flavor groups than usual views.
\end{abstract}


 \maketitle

 \section{Introduction}
\label{sec:intro}

 Recent reactor neutrino  experiments have confirmed a nonzero $\theta_{13}$~\cite{1,2,3,4,5}, which opens the  window to
 determine the Dirac CP violation phase and the octant of $\theta_{23}$. These experiment results also set more stringent constraints on the theoretical models of neutrinos. As a challenge brought by the progress of experiments, few models based on symmetries~\cite{6} can fit the experiment results completely. Perturbations or other phenomenological considerations should be introduced in the typical mixing patterns such as Tri-Bi-Maximal (TBM)~\cite{7,8}, Golden Ratio Mixing (GRM)~\cite{9,10}, Toorop-Feruglio-Hagedorn Mixing (TFHM)~\cite{11,12} and etc.~\cite{13,14,15}. However, even so, the candidate mixing pattern is not unique, we need to determine all viable leptonic mixing  patterns and find differences between their physical origins and predictions. In this article, we extract all viable mixing patterns at $3\sigma$ level of mixing parameters on the basis of various combinations of correlation of elements of mixing matrix. Special correlations  have been  widely discussed in the literature, see Refs.~\cite{16,17,18} for example. We search for more general correlations to construct the leptonic mixing  pattern. The method to generate general correlations is employing the combinations of elementary correlations. We propose two type of elementary correlations. By examining all possible combinations of these elementary correlations, we obtain 62 viable mixing patters, including some familiar types such as Tri-Maximal TM$_{1}$, TM$_{2}$~\cite{19,20,21,22}, $\mu-\tau$ symmetric mixing~\cite{23,24,25,26,45} and many new types. Most of these patterns could be paired via the $\mu-\tau$ interchange which changes the octant of $\theta_{23}$ and the sign of $\cos{\delta}$. Furthermore, according to the dependence of $\cos{\delta}$ on small $\sin^{2}\theta_{13}$ all these patterns can be classified into two types: the perturbative and nonperturbative patterns. The former could be obtained from  perturbing TBM. The latter cannot be obtained from  perturbing any mixing pattern whose $\theta_{13}$ is zero.

 In order to study the robustness of leptonic mixing  models on the basis of flavor groups, the dependence of mass matrix of neutrinos on small mixing parameter $\sin\theta_{13}$ is discussed via special perturbative and nonperturbative mixing patterns. In general cases, a small variation of  $\sin\theta_{13}$ may correspond to large modifications to alignment of vacuum expectation
values in a mixing model on the basis of flavor groups.

The article is organized as follows: In Sec. 2 we introduce the general framework of our program, including the unitary parametrization of mixing matrix, constraints on the parameters at $3\sigma$ level, two types of elementary correlations and the $\mu-\tau$ interchange. In Se. 3 we show all viable combinations of elementary correlations, including the combination of two correlations and the combination of three correlations, mixing angles and Dirac CP phases of viable patterns. In Sec. 4 we discuss evolutions of mass matrices of neutrinos via  special leptonic mixing  patterns. Finally, we present a summary.

\section{General framework}
\label{sec:frame}

\subsection{Unitary parametrization of leptonic mixing  matrix}
In order to simplify calculations of combinations of correlations, we adopt the following parametrization of leptonic mixing  matrix:
\begin{equation}
\label{eq:2.1}
|U|=
\left(
\begin{array}{ccc}
 \sqrt{\frac{-1-k+N}{N}} & \sqrt{\frac{k}{N}} & \sqrt{\frac{1}{N}} \\
 \sqrt{\frac{-l-j+N}{N}} & \sqrt{\frac{l}{N}} & \sqrt{\frac{j}{N}} \\
 \sqrt{\frac{1+l+j+k-N}{N}} & \sqrt{\frac{-l-k+N}{N}} & \sqrt{\frac{-1-j+N}{N}}
\end{array}
\right),
\end{equation}
where\emph{ j, k, l, N} are non-negative real numbers. By comparison with the standard parametrization of PMNS matirx~\cite{27}:
\begin{equation}
\label{eq:2.2}
U=
\left(
\begin{array}{ccc}
 c_{12}c_{13} & s_{12}c_{13} & s_{13}e^{-i\delta} \\
 -s_{12}c_{23}-c_{12}s_{13}s_{23}e^{i\delta} & c_{12}c_{23}-s_{12}s_{13}s_{23}e^{i\delta} & c_{13}s_{23} \\
s_{12}s_{23}-c_{12}s_{13}c_{23}e^{i\delta} & -c_{12}s_{23}-s_{12}s_{13}c_{23}e^{i\delta} & c_{13}c_{23}
\end{array}
\right)
\left(
\begin{array}{ccc}
e^{i\alpha_{1}} & 0 & 0 \\
 0 & e^{i\alpha_{2}} & 0 \\
0 & 0 & 1
\end{array}
\right),
\end{equation}
where $s_{ij}\equiv\sin{\theta_{ij}}$, $c_{ij}\equiv\cos{\theta_{ij}}$, $\delta$ is the Dirac CP-violating phase, $\alpha_{1}$ and $\alpha_{2}$ are Majorana phases. In our parametrization, $s_{ij}$ could be written as:
\begin{equation}
\label{eq:2.3}
\begin{split}
\sin{\theta_{13}} &= \sqrt{\frac{1}{N}} ~,
\\
\sin{\theta_{23}} &= \sqrt{\frac{j}{N-1}} ~,
\\
\sin{\theta_{12}} & = \sqrt{\frac{k}{N-1}} ~.
\end{split}
\end{equation}
 And the Dirac CP phase is expressed as:
\begin{equation}
\label{eq:2.4}
\cos{\delta} = \frac{ -l(-1+N)^2-(1+k-N)(-1+N)N+j(k+N+k N-N^2)}{2\sqrt{N} \sqrt{j(-1-j+N)} \sqrt{k (-1-k+N)}}.
\end{equation}
According to the recent fit data of the Ref.~\cite{28}, absolute values of elements of mixing matrix in $3\sigma$ ranges are:
\begin{equation}
\label{eq:2.5}
|U|=
\left(
\begin{array}{ccc}
 0.801\rightarrow0.845 & 0.514\rightarrow0.580 & 0.137\rightarrow0.158 \\
 0.225\rightarrow0.517 & 0.441-0.699 & 0.614\rightarrow0.793 \\
0.246\rightarrow0.529 &0.464\rightarrow0.713 & 0.590\rightarrow0.776
\end{array}
\right).
\end{equation}
So constraints on our parameters at $3\sigma$ level are:
\begin{equation}
\label{eq:2.6}
\begin{array}{c}
  \frac{1}{N}=  0.0188\rightarrow0.0251, \\
\frac{l}{N}=    0.1945\rightarrow0.4886, \\
 \frac{j}{N}=  0.3770\rightarrow0.6288,\\
\frac{k}{N}=  0.2642\rightarrow0.3364.
\end{array}
\end{equation}

\subsection{Elementary correlations}
\label{sec:element}
 We propose two types of elementary correlations as follows:
\begin{equation}
\label{eq:2.7}
\begin{array}{c}
 A:  ~\ |U_{\alpha i}|= |U_{\beta j}|, \\
B:   ~\ |U_{\alpha i}|= 2|U_{\beta j}|.
\end{array}
\end{equation}
These two types of correlations are implied in some familiar mixing patterns such as TBM and GRM. We examine correlations of these two types. The parameter-space of a viable correlation (or viable combination of elementary correlations) should satisfy two conditions: First, the magnitude of every element of mixing matrix is in $3\sigma$ range; Second, $|\cos{\delta}(l,j,k,N)|\leq1$.
There are 9 viable type-A elementary correlations in total. They are listed as follows:
\begin{equation}
 A1:   ~\ |U_{\mu 1}|= |U_{\tau1}|,
 \end{equation}
 \begin{equation}
 A2:   ~\ |U_{\mu 2}|= |U_{\tau 2}|,
  \end{equation}
 \begin{equation}
 A3:    ~\ |U_{\mu 3}|= |U_{\tau 3}|,
 \end{equation}
 \begin{equation}
 A4:    ~\ |U_{e 2}|= |U_{\mu2}|,
 \end{equation}
 \begin{equation}
 A5:    ~\ |U_{e2}|= |U_{\tau2}|,
 \end{equation}
 \begin{equation}
 A6:    ~\ |U_{\mu2}|= |U_{\mu3}|,
 \end{equation}
 \begin{equation}
 A7:    ~\ |U_{\tau2}|= |U_{\tau3}|,
 \end{equation}
 \begin{equation}
 A8:    ~\ |U_{\mu1}|= |U_{\mu2}|,
 \end{equation}
 \begin{equation}
A9:     ~\ |U_{\tau1}|= |U_{\tau2}|.
\end{equation}
There are 14 viable type-B elementary correlations in total. They are written as follows:
\begin{equation}
 B1:    ~\ |U_{e1}|= 2|U_{\mu 1}|,
 \end{equation}
 \begin{equation}
 B2:    ~\ |U_{e1}|= 2|U_{\tau 1}|,
 \end{equation}
 \begin{equation}
 B3:    ~\ |U_{e2}|= 2|U_{\mu 1}|,
 \end{equation}
 \begin{equation}
 B4:    ~\ |U_{e 2}|= 2|U_{\tau1}|,
 \end{equation}
 \begin{equation}
 B5:    ~\ |U_{\mu2}|= 2|U_{\mu1}|,
 \end{equation}
 \begin{equation}
 B6:    ~\ |U_{\tau2}|= 2|U_{\tau1}|, \end{equation}
 \begin{equation}
 B7:    ~\ |U_{\tau2}|= 2|U_{\mu1}|,
 \end{equation}
 \begin{equation}
 B8:    ~\ |U_{\mu2}|= 2|U_{\tau1}|,
 \end{equation}
 \begin{equation}
B9:     ~\ |U_{\mu3}|= 2|U_{\mu1}|,
\end{equation}
 \begin{equation}
B10:     ~\ |U_{\tau3}|= 2|U_{\tau1}|,
\end{equation}
 \begin{equation}
B11:     ~\ |U_{\mu3}|= 2|U_{\tau1}|,
\end{equation}
 \begin{equation}
B12:     ~\ |U_{\tau3}|= 2|U_{\mu1}|,
\end{equation}
 \begin{equation}
B13:     ~\ |U_{\mu1}|= 2|U_{\tau1}|,
\end{equation}
 \begin{equation}
B14:     ~\ |U_{\tau1}|= 2|U_{\mu1}|.
\end{equation}
In order to get general mixing patterns, all combinations of elementary correlations are examined. The number of possible combinations is so large. However, because the octant of $\theta_{23}$ and CP violation phase $\delta$ are uncertain at $3\sigma$ level, one type of combination can be converted to another type through the $\mu-\tau$ interchange. $\mu-\tau$ interchange is equivalent to the following transformation:
\begin{equation}
\label{eq:2.10}
\left \{{
\begin{array}{c}
  \theta_{23}\rightarrow\frac{\pi}{2}- \theta_{23}\\
\delta\rightarrow\delta+\pi
\end{array}}
\right.
~~~~or~~~~~ \left \{{
\begin{array}{c}
  j \rightarrow N-1-j\\
l \rightarrow N-l-k .
\end{array}}
\right.
\end{equation}
As for type-A correlations, $\mu-\tau$ interchange brings following correspondences:
\begin{equation}
\label{eq:2.11}
\begin{array}{c}
 A1 \leftrightarrow A1, ~~ A2 \leftrightarrow A2, ~~ A3 \leftrightarrow A3,\\
   A4 \leftrightarrow A5, ~~ A6 \leftrightarrow A7, ~~ A8 \leftrightarrow A9.
\end{array}
\end{equation}
And correspondences in type-B correlations  are:
\begin{equation}
\label{eq:2.12}
\begin{array}{c}
 B1 \leftrightarrow B2, ~~ B3 \leftrightarrow B4, ~~ B5 \leftrightarrow B6, ~~B7 \leftrightarrow B8,\\
    B9\leftrightarrow B10, ~~B11 \leftrightarrow B12, ~~ B13 \leftrightarrow B14.
\end{array}
\end{equation}
So we first examine the parameter-space of a combination of elementary correlations. Then the examination of the corresponding combination is carried out   easily.
\section{Combinations of elementary correlations}
\label{sec:patterns}
\subsection{Viable combinations of two elementary correlations}
\label{sec:two}
There are 83 viable combinations of two elementary correlations in total. They can be classified into 3 types: Ai-Aj, Bi-Bj, and Ai-Bj, where the notation Ai-Aj means elementary correlation Ai and Aj hold at the same time. The common combinations of these three types are not counted repeatedly.

There are 12 viable Ai-Aj combinations. They are listed as follows:
\begin{equation}
\label{eq:3.1}
\begin{array}{c}
A1-A2, ~~A4-A5, ~~A1-A4, ~~A1-A5, ~~A1-A6,  ~~A1-A7, \\
A3-A4,~~A3-A5, ~~A5-A6, ~~A4-A7, ~~A7-A8, ~~~A6-A9.
\end{array}
\end{equation}
There are 23 viable Bi-Bj combinations. They are listed as follows:
\begin{equation}
\label{eq:3.2}
\begin{array}{c}
B1-B2, ~~B1-B6, ~~B2-B5, ~~B1-B8,  ~~B2-B7, ~~B1-B10,\\
B2-B9, ~~B1-B11, ~~B2-B12, ~~B3-B5, ~~B4-B6, ~~B3-B7, \\
B4-B8, ~~B5-B7, ~~B6-B8, ~~B5-B9,  ~~B6-B10, ~~B7-B12,  \\
B8-B11, ~~B9-B11, ~~B10-B12, ~~B9-B12, ~~B10-B11.
\end{array}
\end{equation}
There are 48 viable Ai-Bj combinations. They are listed as follows:
\begin{equation}
\label{eq:3.3}
\begin{array}{c}
A4-B14, ~~A8-B2, ~~A2-B2, ~~A2-B1, ~~A2-B3, ~~A2-B4,\\
A2-B9, ~~A2-B10, ~~A2-B11, ~~A2-B12, ~~A2-B13, ~~A2-B14, \\
A3-B2, ~~A3-B1, ~~A3-B5, ~~A3-B6, ~~A4-B2, ~~A5-B1,\\
A4-B1, ~~A5-B2, ~~ A4-B6, ~~A5-B5, ~~A4-B7, ~~A5-B8,\\
A4-B9, ~~A5-B10, ~~A4-B10, ~~A5-B9, ~~A4-B11, ~~A5-B12,\\
A4-B12, ~~A5-B11, ~~A6-B2, ~~A7-B1, ~~A6-B1,~~A7-B2,\\
A6-B4, ~~A7-B3, ~~A6-B6, ~~A7-B5, ~~A6-B10, ~~A7-B9,\\
A6-B12, ~~A7-B11,~~A6-B13, ~~A7-B14, ~A8-B11, ~~A9-B12.
\end{array}
\end{equation}
Let us describe these viable combinations in more detail:\\
$\bullet$  ~~Most viable combinations are paired via the $\mu-\tau$ interchange. There are only 5 single combinations, among which 3 combinations are invariant under the $\mu-\tau$ interchange, including A1-A2 ($\mu-\tau$ symmetry), B1-B2 (TM$_{1}$), A4-A5 (TM$_{2}$). The reason why A8-B2 and A4-B14 are single is that the constraint matrix \eqref{eq:2.5} is not symmetric under the $\mu-\tau$ interchange. So the viable parameter-space of a combination  may become unviable under the $\mu-\tau$ interchange, especially when the parameter-space is small.\\
$\bullet$  ~~For more than half of the viable combinations, their 2-dimensional parameter-spaces are not compressed obviously. There are 21 combinations whose parameter-spaces are reduced more than half compared with the non-correlation case. These combinations are: A4-A5, A7-A8, A6-A9, B1-B8,
B2-B7, B1-B7, B2-B5, B3-B7, B4-B8, B4-B6, B9-B11, B10-B12, A2-B13, A2-B14, A4-B8, A4-B14, A5-B11, A6-B13, A8-B2, A8-B11, A9-B12.\\
$\bullet$  ~~There are 17 combinations, whose octants of $\theta_{23}$ are uncertain. These combinations are: A4-A5, B1-B2, B1-B10, B1-B11, B2-B5, B2-B9, B2-B12,
A2-B2, A2-B1, A4-B2, A4-B6, A4-B10, A4-B11, A5-B1, A5-B5, A5-B9, A5-B12.

\subsection{Viable combinations of three elementary correlations}
\label{sec:three}
There are 62 viable combinations of three elementary correlations in total. They can be classified into 4 types: Ai-Aj-Ak, Bi-Bj-Bk, Ai-Bj-Bk, and Ai-Aj-Bk. The common combinations of these four types are not counted repeatedly. In detail, there are 3 viable Ai-Aj-Ak combinations. Their mixing patterns are listed as follows:
 \begin{equation}
 \label{eq:3.4}
A1-A4-A7: ~~ |U| = \left(
\begin{array}{ccc}
 \sqrt{\frac{5-8 s}{7}} & \sqrt{\frac{2+s}{7}} & \sqrt{s} \\
 \sqrt{\frac{1+4 s}{7}} & \sqrt{\frac{2+s}{7}} & \sqrt{\frac{4-5s}{7}} \\
\sqrt{\frac{1+4 s}{7}} &\sqrt{\frac{3-2s}{7}} &\sqrt{\frac{3-2s}{7}}
\end{array}
\right),
\end{equation}

 \begin{equation}
A1-A5-A6: ~~ |U|= \left(
\begin{array}{ccc}
\sqrt{\frac{5-8 s}{7}} & \sqrt{\frac{2+s}{7}} & \sqrt{s} \\
\sqrt{\frac{1+4 s}{7}} &\sqrt{\frac{3-2s}{7}} &\sqrt{\frac{3-2s}{7}}\\
 \sqrt{\frac{1+4 s}{7}} & \sqrt{\frac{2+s}{7}} & \sqrt{\frac{4-5s}{7}}
\end{array}
\right),
\end{equation}

 \begin{equation}
 \label{eq:3.6}
A1-A4-A5: ~~ |U| = \left(
\begin{array}{ccc}
 \sqrt{\frac{2-3s}{3}} & \sqrt{\frac{1}{3}} & \sqrt{s} \\
 \sqrt{\frac{1+3 s}{6}} & \sqrt{\frac{1}{3}} & \sqrt{\frac{1-s}{2}} \\
 \sqrt{\frac{1+3 s}{6}} & \sqrt{\frac{1}{3}} & \sqrt{\frac{1-s}{2}}
\end{array}
\right),
\end{equation}
where $s=\frac{1}{N}=\sin^{2}{\theta_{13}}$.

There are 5 viable Bi-Bj-Bk combinations. Their mixing patterns are listed as follows:
\begin{equation}
B1-B10-B11: ~~ |U| = \left(
\begin{array}{ccc}
\sqrt{\frac{7+s}{10}} & \sqrt{\frac{3-11 s}{10}} & \sqrt{s} \\
 \sqrt{\frac{7+s}{40}} & \sqrt{\frac{13+19s}{40}} & \sqrt{\frac{1-s}{2}} \\
 \sqrt{\frac{1-s}{8}} & \sqrt{\frac{3+5s}{8}} & \sqrt{\frac{1-s}{2}}
\end{array}
\right),
\end{equation}

\begin{equation}
B2-B9-B12: ~~ |U| =
\left(
\begin{array}{ccc}
\sqrt{\frac{7+s}{10}} & \sqrt{\frac{3-11 s}{10}} & \sqrt{s} \\
 \sqrt{\frac{1-s}{8}} & \sqrt{\frac{3+5s}{8}} & \sqrt{\frac{1-s}{2}}\\
 \sqrt{\frac{7+s}{40}} & \sqrt{\frac{13+19s}{40}} & \sqrt{\frac{1-s}{2}}
\end{array}
\right),
\end{equation}

\begin{equation}
B1-B6-B10: ~~ |U| = \left(
\begin{array}{ccc}
 \frac{4 }{3}\sqrt{\frac{2}{5}} & \sqrt{\frac{13}{45}-s} & \sqrt{s} \\
 \frac{2 }{3}\sqrt{\frac{2}{5}} & \sqrt{\frac{4}{15}+s} & \frac{1}{3} \sqrt{5-9 s} \\
 \frac{1}{3} & \frac{2}{3} & \frac{2}{3}
\end{array}
\right),
\end{equation}

\begin{equation}
B2-B5-B9: ~~ |U| =\left(
\begin{array}{ccc}
 \frac{4 }{3}\sqrt{\frac{2}{5}} & \sqrt{\frac{13}{45}-s} & \sqrt{s} \\
 \frac{1}{3} & \frac{2}{3} & \frac{2}{3} \\
 \frac{2 }{3}\sqrt{\frac{2}{5}} & \sqrt{\frac{4}{15}+s} & \frac{1}{3} \sqrt{5-9 s}
\end{array}
\right),
\end{equation}

\begin{equation}
B3-B5-B7: ~~ |U|= \left(
\begin{array}{ccc}
 \sqrt{\frac{2}{3}-s} & \sqrt{\frac{1}{3}} & \sqrt{s} \\
 \frac{1}{2}\sqrt{\frac{1}{3}} & \sqrt{\frac{1}{3}} & \frac{1}{2}\sqrt{\frac{7}{3}} \\
 \sqrt{\frac{1}{4}+s} & \sqrt{\frac{1}{3}} & \sqrt{\frac{5}{12}-s}
\end{array}
\right).
\end{equation}

There are 35 viable Ai-Bj-Bk combinations. Their mixing patterns are listed as follows:
\begin{equation}
A4-B1-B2: ~~ |U| = \left(
\begin{array}{ccc}
 \sqrt{\frac{2}{3}} & \sqrt{\frac{1}{3}-s} & \sqrt{s} \\
 \sqrt{\frac{1}{6}} & \sqrt{\frac{1}{3}-s} & \sqrt{\frac{1}{2}+s} \\
 \sqrt{\frac{1}{6}} & \sqrt{\frac{1}{3}+2 s} & \sqrt{\frac{1}{2}-2 s}
\end{array}
\right),
\end{equation}

\begin{equation}
A5-B1-B2: ~~ |U| = \left(
\begin{array}{ccc}
 \sqrt{\frac{2}{3}} & \sqrt{\frac{1}{3}-s} & \sqrt{s} \\
 \sqrt{\frac{1}{6}} & \sqrt{\frac{1}{3}+2 s} & \sqrt{\frac{1}{2}-2 s}\\
 \sqrt{\frac{1}{6}} & \sqrt{\frac{1}{3}-s} & \sqrt{\frac{1}{2}+s}
 \end{array}
\right)
\end{equation}

\begin{equation}
~\label{eq:3.14}
A4-B1-B6: ~~ |U|= \left(
\begin{array}{ccc}
 \sqrt{\frac{5-2 s}{7}} & \sqrt{\frac{2-5s}{7}} & \sqrt{s} \\
 \sqrt{\frac{5-2 s}{28}} & \sqrt{\frac{2-5s}{7}} & \sqrt{\frac{15+22 s}{28}} \\
 \sqrt{\frac{3+10 s}{28}} & \sqrt{\frac{3+10 s}{7}} & \sqrt{\frac{13-50 s}{28}}
\end{array}
\right),
\end{equation}

\begin{equation}
A5-B2-B5: ~~ |U| = \left(
\begin{array}{ccc}
 \sqrt{\frac{5-2 s}{7}} & \sqrt{\frac{2-5s}{7}} & \sqrt{s} \\
 \sqrt{\frac{3+10 s}{28}} & \sqrt{\frac{3+10 s}{7}} & \sqrt{\frac{13-50 s}{28}}\\
 \sqrt{\frac{5-2 s}{28}} & \sqrt{\frac{2-5s}{7}} & \sqrt{\frac{15+22 s}{28}}
 \end{array}
\right),
\end{equation}

\begin{equation}
A4-B1-B10: ~~ |U| =\left(
\begin{array}{ccc}
 2\sqrt{\frac{3+2 s}{17}} & \sqrt{\frac{5-25s}{17}}  & \sqrt{s} \\
 \sqrt{\frac{3+2 s}{17}}  & \sqrt{\frac{5-25s}{17}}  & \sqrt{\frac{9+23 s}{17}} \\
 \sqrt{\frac{2-10s}{17}}  & \sqrt{\frac{7+50 s}{17}}  & 2 \sqrt{\frac{2-10s}{17}}
\end{array}
\right),
\end{equation}

\begin{equation}
A5-B2-B9: ~~ |U|= \left(
\begin{array}{ccc}
 2\sqrt{\frac{3+2 s}{17}} & \sqrt{\frac{5-25s}{17}}  & \sqrt{s} \\
  \sqrt{\frac{2-10s}{17}}  & \sqrt{\frac{7+50 s}{17}}  & 2 \sqrt{\frac{2-10s}{17}}\\
 \sqrt{\frac{3+2 s}{17}}  & \sqrt{\frac{5-25s}{17}}  & \sqrt{\frac{9+23 s}{17}}
 \end{array}
\right),
\end{equation}

\begin{equation}
A4-B7-B12: ~~ |U|=\left(
\begin{array}{ccc}
 \sqrt{\frac{7-14s}{10}}  & \sqrt{\frac{3+4s}{10}} & \sqrt{s} \\
 \sqrt{\frac{1-2s}{10}}  & \sqrt{\frac{3+4s}{10}} & \sqrt{\frac{3-s}{5}} \\
 \sqrt{\frac{1+8s}{5}} & \sqrt{\frac{2-4s}{5}}  & \sqrt{\frac{2-4s}{5}}
\end{array}
\right),
\end{equation}

\begin{equation}
A5-B8-B11: ~~ |U|=\left(
\begin{array}{ccc}
 \sqrt{\frac{7-14s}{10}}  & \sqrt{\frac{3+4s}{10}} & \sqrt{s} \\
 \sqrt{\frac{1+8s}{5}} & \sqrt{\frac{2-4s}{5}}  & \sqrt{\frac{2-4s}{5}}\\
 \sqrt{\frac{1-2s}{10}}  & \sqrt{\frac{3+4s}{10}} & \sqrt{\frac{3-s}{5}}
 \end{array}
\right),
\end{equation}

\begin{equation}
A4-B6-B10: ~~ |U| =  \left(
\begin{array}{ccc}
 \sqrt{\frac{13}{18}-s} & \frac{1}{3}\sqrt{\frac{5}{2}} & \sqrt{s} \\
 \sqrt{\frac{1}{6}+s} & \frac{1}{3}\sqrt{\frac{5}{2}} & \frac{1}{3} \sqrt{5-9 s} \\
 \frac{1}{3} & \frac{2}{3} & \frac{2}{3}
\end{array}
\right),
\end{equation}

\begin{equation}
A5-B5-B9: ~~ |U|  = \left(
\begin{array}{ccc}
 \sqrt{\frac{13}{18}-s} & \frac{1}{3}\sqrt{\frac{5}{2}} & \sqrt{s} \\
  \frac{1}{3} & \frac{2}{3} & \frac{2}{3}\\
 \sqrt{\frac{1}{6}+s} & \frac{1}{3}\sqrt{\frac{5}{2}} & \frac{1}{3} \sqrt{5-9 s}
 \end{array}
\right),
\end{equation}

\begin{equation}
A4-B9-B11: ~~ |U|  =
\left(
\begin{array}{ccc}
 \sqrt{\frac{5-2s}{7}} & \sqrt{\frac{2-5s}{7}} & \sqrt{s} \\
 \sqrt{\frac{1+s}{7}} & \sqrt{\frac{2-5s}{7}} & 2\sqrt{\frac{1+s}{7}} \\
 \sqrt{\frac{1+s}{7}} & \sqrt{\frac{3+10s}{7}} & \sqrt{\frac{3-11s}{7}}
\end{array}
\right),
\end{equation}

\begin{equation}
A5-B10-B12: ~~ |U| =
\left(
\begin{array}{ccc}
 \sqrt{\frac{5-2s}{7}} & \sqrt{\frac{2-5s}{7}} & \sqrt{s} \\
 \sqrt{\frac{1+s}{7}} & \sqrt{\frac{3+10s}{7}} & \sqrt{\frac{3-11s}{7}}\\
 \sqrt{\frac{1+s}{7}} & \sqrt{\frac{2-5s}{7}} & 2\sqrt{\frac{1+s}{7}}
 \end{array}
\right),
\end{equation}

\begin{equation}
A4-B1-B11: ~~ |U|  =
\left(
\begin{array}{ccc}
 2\sqrt{\frac{4-s}{23}} & \sqrt{\frac{7-19s}{23}} & \sqrt{s} \\
 \sqrt{\frac{4-s}{23}} & \sqrt{\frac{7-19s}{23}} & 2\sqrt{\frac{3+5s}{23}} \\
 \sqrt{\frac{3+5s}{23}} & \sqrt{\frac{9+38s}{23}} & \sqrt{\frac{11-43s}{23}}
\end{array}
\right),
\end{equation}

\begin{equation}
A5-B2-B12: ~~ |U| =
\left(
\begin{array}{ccc}
 2\sqrt{\frac{4-s}{23}} & \sqrt{\frac{7-19s}{23}} & \sqrt{s} \\
 \sqrt{\frac{3+5s}{23}} & \sqrt{\frac{9+38s}{23}} & \sqrt{\frac{11-43s}{23}}\\
 \sqrt{\frac{4-s}{23}} & \sqrt{\frac{7-19s}{23}} & 2\sqrt{\frac{3+5s}{23}}
 \end{array}
\right),
\end{equation}

\begin{equation}
\label{eq:3.26}
A4-B2-B7: ~~ |U| =
\left(
\begin{array}{ccc}
 \sqrt{\frac{5-2s}{7}} & \sqrt{\frac{2-5s}{7}} & \sqrt{s} \\
 \sqrt{\frac{3+10s}{28}} & \sqrt{\frac{2-5s}{7}} & \sqrt{\frac{17+10s}{28}} \\
 \sqrt{\frac{5-2s}{28}} & \sqrt{\frac{3+10s}{7}} & \sqrt{\frac{11-38s}{28}}
\end{array}
\right),
\end{equation}

\begin{equation}
A4-B2-B9: ~~ |U| =
\left(
\begin{array}{ccc}
 2\sqrt{\frac{5-s}{29}} & \sqrt{\frac{9-25s}{29}} & \sqrt{s} \\
 \sqrt{\frac{4+5s}{29}} & \sqrt{\frac{9-25s}{29}} & 2\sqrt{\frac{4+5s}{29}} \\
 \sqrt{\frac{5-s}{29}} & \sqrt{\frac{11+50s}{29}} & \sqrt{\frac{13-49s}{29}}
\end{array}
\right),
\end{equation}

\begin{equation}
A5-B1-B10: ~~ |U| =
\left(
\begin{array}{ccc}
 2\sqrt{\frac{5-s}{29}} & \sqrt{\frac{9-25s}{29}} & \sqrt{s} \\
 \sqrt{\frac{5-s}{29}} & \sqrt{\frac{11+50s}{29}} & \sqrt{\frac{13-49s}{29}}\\
 \sqrt{\frac{4+5s}{29}} & \sqrt{\frac{9-25s}{29}} & 2\sqrt{\frac{4+5s}{29}}
 \end{array}
\right),
\end{equation}

\begin{equation}
A4-B10-B11: ~~ |U|  =
\left(
\begin{array}{ccc}
 \frac{1}{4} \sqrt{11-11s} & \frac{1}{4} \sqrt{5-5 s} & \sqrt{s} \\
 \frac{1}{4} \sqrt{3+13 s} & \frac{1}{4} \sqrt{5-5 s} & \sqrt{\frac{1-s}{2}} \\
 \sqrt{\frac{1-s}{8}} & \sqrt{\frac{3+5s}{8}} & \sqrt{\frac{1-s}{2}}
\end{array}
\right),
\end{equation}

\begin{equation}
\label{eq:3.30}
A5-B9-B12: ~~ |U|  =
\left(
\begin{array}{ccc}
 \frac{1}{4} \sqrt{11-11s} & \frac{1}{4} \sqrt{5-5 s} & \sqrt{s} \\
 \sqrt{\frac{1-s}{8}} & \sqrt{\frac{3+5s}{8}} & \sqrt{\frac{1-s}{2}}\\
 \frac{1}{4} \sqrt{3+13 s} & \frac{1}{4} \sqrt{5-5 s} & \sqrt{\frac{1-s}{2}} \
 \end{array}
\right),
\end{equation}

\begin{equation}
\label{eq:3.31}
A2-B1-B2: ~~ |U|  =
\left(
\begin{array}{ccc}
 \sqrt{\frac{2}{3}} & \sqrt{\frac{1}{3}-s} & \sqrt{s} \\
 \sqrt{\frac{1}{6}} & \sqrt{\frac{2+3s}{6}} & \sqrt{\frac{1-s}{2}} \\
 \sqrt{\frac{1}{6}} & \sqrt{\frac{2+3s}{6}} & \sqrt{\frac{1-s}{2}}
\end{array}
\right),
\end{equation}

\begin{equation}
A2-B1-B10: ~~ |U| =
\left(
\begin{array}{ccc}
 2\sqrt{\frac{8+s}{46}}  & \sqrt{\frac{7-25s}{23}} & \sqrt{s} \\
 \sqrt{\frac{8+s}{46}} & \sqrt{\frac{16+25s}{46}} & \sqrt{\frac{11-13s}{23}} \\
 \sqrt{\frac{6-5s}{46}} & \sqrt{\frac{16+25s}{46}} & 2\sqrt{\frac{6-5s}{46}}
\end{array}
\right),
\end{equation}

\begin{equation}
A2-B2-B9: ~~ |U|  =
\left(
\begin{array}{ccc}
 2\sqrt{\frac{8+s}{46}}  & \sqrt{\frac{7-25s}{23}} & \sqrt{s} \\
  \sqrt{\frac{6-5s}{46}} & \sqrt{\frac{16+25s}{46}} & 2\sqrt{\frac{6-5s}{46}}\\
 \sqrt{\frac{8+s}{46}} & \sqrt{\frac{16+25s}{46}} & \sqrt{\frac{11-13s}{23}}
\end{array}
\right),
\end{equation}

\begin{equation}
A2-B1-B11: ~~ |U|  =
\left(
\begin{array}{ccc}
 2\sqrt{\frac{6+s}{34}} & \sqrt{\frac{5-19s}{17}} & \sqrt{s} \\
 \sqrt{\frac{6+s}{34}} & \sqrt{\frac{12+19s}{17}} & 2\sqrt{\frac{4-5s}{34}} \\
 \sqrt{\frac{4-5s}{34}} & \sqrt{\frac{12+19s}{17}} & \sqrt{\frac{9-7s}{17}}
\end{array}
\right),
\end{equation}

\begin{equation}
A2-B2-B12: ~~ |U|  =
\left(
\begin{array}{ccc}
 2\sqrt{\frac{6+s}{34}} & \sqrt{\frac{5-19s}{17}} & \sqrt{s} \\
  \sqrt{\frac{4-5s}{34}} & \sqrt{\frac{12+19s}{17}} & \sqrt{\frac{9-7s}{17}}\\
 \sqrt{\frac{6+s}{34}} & \sqrt{\frac{12+19s}{17}} & 2\sqrt{\frac{4-5s}{34}}
\end{array}
\right),
\end{equation}

\begin{equation}
A3-B1-B10: ~~ |U|  =
\left(
\begin{array}{ccc}
 2\sqrt{\frac{7+s}{40}} & \sqrt{\frac{3-11s}{10}} & \sqrt{s} \\
 \sqrt{\frac{7+s}{40}} & \sqrt{\frac{13+19s}{40}} & \sqrt{\frac{1-s}{2}} \\
 \sqrt{\frac{1-s}{8}} & \sqrt{\frac{3+5s}{8}} & \sqrt{\frac{1-s}{2}}
\end{array}
\right),
\end{equation}

\begin{equation}
A3-B2-B9: ~~ |U|  =
\left(
\begin{array}{ccc}
 2\sqrt{\frac{7+s}{40}} & \sqrt{\frac{3-11s}{10}} & \sqrt{s} \\
 \sqrt{\frac{1-s}{8}} & \sqrt{\frac{3+5s}{8}} & \sqrt{\frac{1-s}{2}}\\
 \sqrt{\frac{7+s}{40}} & \sqrt{\frac{13+19s}{40}} & \sqrt{\frac{1-s}{2}}
 \end{array}
\right),
\end{equation}

\begin{equation}
A7-B1-B2: ~~ |U|  =
\left(
\begin{array}{ccc}
 \sqrt{\frac{2}{3}} & \sqrt{\frac{1}{3}-s} & \sqrt{s} \\
 \sqrt{\frac{1}{6}} & \frac{1}{2} \sqrt{1+4 s} & \sqrt{\frac{7}{12}-s} \\
 \sqrt{\frac{1}{6}} & \sqrt{\frac{5}{12}} & \sqrt{\frac{5}{12}}
\end{array}
\right),
\end{equation}

\begin{equation}
A6-B1-B2: ~~ |U| =
\left(
\begin{array}{ccc}
 \sqrt{\frac{2}{3}} & \sqrt{\frac{1}{3}-s} & \sqrt{s} \\
  \sqrt{\frac{1}{6}} & \sqrt{\frac{5}{12}} & \sqrt{\frac{5}{12}}\\
 \sqrt{\frac{1}{6}} & \frac{1}{2} \sqrt{1+4 s} & \sqrt{\frac{7}{12}-s}
\end{array}
\right),
\end{equation}

\begin{equation}
A7-B1-B6: ~~ |U|  =
\left(
\begin{array}{ccc}
 \frac{4 }{3}\sqrt{\frac{2}{5}} & \sqrt{\frac{13}{45}-s} & \sqrt{s} \\
 \frac{2 }{3}\sqrt{\frac{2}{5}} & \sqrt{\frac{4}{15}+s} & \frac{1}{3} \sqrt{5-9 s} \\
 \frac{1}{3} & \frac{2}{3} & \frac{2}{3}
\end{array}
\right),
\end{equation}

\begin{equation}
A6-B2-B5: ~~ |U|  =
\left(
\begin{array}{ccc}
 \frac{4 }{3}\sqrt{\frac{2}{5}} & \sqrt{\frac{13}{45}-s} & \sqrt{s} \\
  \frac{1}{3} & \frac{2}{3} & \frac{2}{3}\\
 \frac{2 }{3}\sqrt{\frac{2}{5}} & \sqrt{\frac{4}{15}+s} & \frac{1}{3} \sqrt{5-9 s}
\end{array}
\right),
\end{equation}

\begin{equation}
A7-B1-B11: ~~ |U|  =
\left(
\begin{array}{ccc}
 2 \sqrt{\frac{6+2s}{35}}  & \sqrt{\frac{11-43s}{35}}  & \sqrt{s} \\
 \sqrt{\frac{6+2s}{35}}  & \sqrt{\frac{9+38s}{35}}  & \sqrt{\frac{4-8s}{7}} \\
 \sqrt{\frac{1-2s}{7}} & \sqrt{\frac{3+s}{7}}  & \sqrt{\frac{3+s}{7}}
\end{array}
\right),
\end{equation}

\begin{equation}
A6-B2-B12: ~~ |U| =
\left(
\begin{array}{ccc}
 2 \sqrt{\frac{6+2s}{35}}  & \sqrt{\frac{11-43s}{35}}  & \sqrt{s} \\
 \sqrt{\frac{1-2s}{7}} & \sqrt{\frac{3+s}{7}}  & \sqrt{\frac{3+s}{7}}\\
 \sqrt{\frac{6+2s}{35}}  & \sqrt{\frac{9+38s}{35}}  & \sqrt{\frac{4-8s}{7}}
 \end{array}
\right),
\end{equation}

\begin{equation}
\label{eq:3.44}
A7-B2-B9: ~~ |U| =
\left(
\begin{array}{ccc}
 2\sqrt{\frac{7+2s}{41}} & \sqrt{\frac{13-49s}{41}} & \sqrt{s} \\
 \sqrt{\frac{6-10s}{41}}  & \sqrt{\frac{11+50s}{41}} & 2 \sqrt{\frac{6-10s}{41}} \\
 \sqrt{\frac{7+2s}{41}} & \sqrt{\frac{17-s}{41}} & \sqrt{\frac{17-s}{41}}
\end{array}
\right),
\end{equation}

\begin{equation}
A6-B1-B10: ~~ |U|  =
\left(
\begin{array}{ccc}
 2\sqrt{\frac{7+2s}{41}} & \sqrt{\frac{13-49s}{41}} & \sqrt{s} \\
  \sqrt{\frac{7+2s}{41}} & \sqrt{\frac{17-s}{41}} & \sqrt{\frac{17-s}{41}}\\
 \sqrt{\frac{6-10s}{41}}  & \sqrt{\frac{11+50s}{41}} & 2 \sqrt{\frac{6-10s}{41}}
\end{array}
\right),
\end{equation}

\begin{equation}
A7-B3-B5: ~~ |U|  =
\left(
\begin{array}{ccc}
 \sqrt{\frac{9-17s}{13}} & 2\sqrt{\frac{1+s}{13}} & \sqrt{s} \\
 \sqrt{\frac{1+s}{13}} & 2\sqrt{\frac{1+s}{13}} & \sqrt{\frac{8-5s}{13}} \\
 \sqrt{\frac{3+16s}{13}} & \sqrt{\frac{5-8s}{13}} & \sqrt{\frac{5-8s}{13}}
\end{array}
\right).
\end{equation}

There are 19 viable Ai-Aj-Bk combinations. Their mixing patterns  are listed as follows:
\begin{equation}
A7-A4-B1: ~~ |U| =
\left(
\begin{array}{ccc}
 2 \sqrt{\frac{2-4s}{11}}  &  \sqrt{\frac{3+5s}{11}}  & \sqrt{s} \\
 \sqrt{\frac{2-4s}{11}}  & \sqrt{\frac{3+5s}{11}}  & \sqrt{\frac{6-s}{11}} \\
 \sqrt{\frac{1+20s}{11}}  & \sqrt{\frac{5-10s}{11}}  & \sqrt{\frac{5-10s}{11}}
\end{array}
\right),
\end{equation}

\begin{equation}
A6-A5-B2: ~~ |U|  =
\left(
\begin{array}{ccc}
 2 \sqrt{\frac{2-4s}{11}}  &  \sqrt{\frac{3+5s}{11}}  & \sqrt{s} \\
 \sqrt{\frac{1+20s}{11}}  & \sqrt{\frac{5-10s}{11}}  & \sqrt{\frac{5-10s}{11}}\\
 \sqrt{\frac{2-4s}{11}}  & \sqrt{\frac{3+5s}{11}}  & \sqrt{\frac{6-s}{11}}
 \end{array}
\right),
\end{equation}

\begin{equation}
A7-A4-B2: ~~ |U|  =
\left(
\begin{array}{ccc}
 2\sqrt{\frac{3-4s}{17}} & \sqrt{\frac{5-s}{17}} & \sqrt{s} \\
 \sqrt{\frac{2+20s}{17}}  & \sqrt{\frac{5-s}{17}} & \sqrt{\frac{10-19s}{17}} \\
 \sqrt{\frac{3-4s}{17}} & \sqrt{\frac{7+2s}{17}} & \sqrt{\frac{7+2s}{17}}
\end{array}
\right),
\end{equation}

\begin{equation}
A6-A5-B1: ~~ |U|  =
\left(
\begin{array}{ccc}
 2\sqrt{\frac{3-4s}{17}} & \sqrt{\frac{5-s}{17}} & \sqrt{s} \\
  \sqrt{\frac{3-4s}{17}} & \sqrt{\frac{7+2s}{17}} & \sqrt{\frac{7+2s}{17}}\\
 \sqrt{\frac{2+20s}{17}}  & \sqrt{\frac{5-s}{17}} & \sqrt{\frac{10-19s}{17}}
\end{array}
\right),
\end{equation}

\begin{equation}
A7-A4-B9: ~~ |U| =
\left(
\begin{array}{ccc}
 \sqrt{\frac{10-19s}{14}} & \sqrt{\frac{4+5s}{14}} & \sqrt{s} \\
 \sqrt{\frac{2-s}{14}} & \sqrt{\frac{4+5s}{14}} & 2\sqrt{\frac{2-s}{14}} \\
 \sqrt{\frac{1+10s}{7}} & \sqrt{\frac{3-5s}{7}} & \sqrt{\frac{3-5s}{7}}
\end{array}
\right),
\end{equation}

\begin{equation}
\label{eq:3.52}
A6-A5-B10: ~~ |U| =
\left(
\begin{array}{ccc}
 \sqrt{\frac{10-19s}{14}} & \sqrt{\frac{4+5s}{14}} & \sqrt{s} \\
 \sqrt{\frac{1+10s}{7}} & \sqrt{\frac{3-5s}{7}} & \sqrt{\frac{3-5s}{7}}\\
 \sqrt{\frac{2-s}{14}} & \sqrt{\frac{4+5s}{14}} & 2\sqrt{\frac{2-s}{14}}
 \end{array}
\right),
\end{equation}

\begin{equation}
A7-A4-B11: ~~ |U|  =
\left(
\begin{array}{ccc}
 \sqrt{\frac{10-13s}{14}} & \sqrt{\frac{4-s}{14}} & \sqrt{s} \\
 \sqrt{\frac{2+17s}{14}} & \sqrt{\frac{4-s}{14}} & 2\sqrt{\frac{1-2s}{7}} \\
 \sqrt{\frac{1-2s}{7}} & \sqrt{\frac{3+s}{7}} & \sqrt{\frac{3+s}{7}}
\end{array}
\right),
\end{equation}

\begin{equation}
A6-A5-B12: ~~ |U| =
\left(
\begin{array}{ccc}
 \sqrt{\frac{10-13s}{14}} & \sqrt{\frac{4-s}{14}} & \sqrt{s} \\
 \sqrt{\frac{1-2s}{7}} & \sqrt{\frac{3+s}{7}} & \sqrt{\frac{3+s}{7}}\\
 \sqrt{\frac{2+17s}{14}} & \sqrt{\frac{4-s}{14}} & 2\sqrt{\frac{1-2s}{7}}
 \end{array}
\right),
\end{equation}

\begin{equation}
A7-A4-B14: ~~ |U|  =
\left(
\begin{array}{ccc}
 \frac{1}{4} \sqrt{11-20 s} & \frac{1}{4} \sqrt{5+4 s} & \sqrt{s} \\
 \frac{1}{4} \sqrt{1+4 s} & \frac{1}{4} \sqrt{5+4 s} & \sqrt{\frac{5}{8}-\frac{s}{2}} \\
 \frac{1}{2} \sqrt{1+4 s} & \frac{1}{2} \sqrt{\frac{3}{2}-2 s} & \sqrt{\frac{3}{8}-\frac{s}{2}}
\end{array}
\right),
\end{equation}

\begin{equation}
A4-A5-B1: ~~ |U|  =
\left(
\begin{array}{ccc}
 \sqrt{\frac{2}{3}-s} & \sqrt{\frac{1}{3}} & \sqrt{s} \\
 \sqrt{\frac{1}{6}-\frac{s}{4}} & \sqrt{\frac{1}{3}} & \frac{\sqrt{2+s}}{2} \\
 \sqrt{\frac{1}{6}+\frac{5 s}{4}} & \sqrt{\frac{1}{3}} & \frac{1}{2} \sqrt{2-5 s}
\end{array}
\right),
\end{equation}

\begin{equation}
A4-A5-B2: ~~ |U|  =
\left(
\begin{array}{ccc}
 \sqrt{\frac{2}{3}-s} & \sqrt{\frac{1}{3}} & \sqrt{s} \\
 \sqrt{\frac{1}{6}+\frac{5 s}{4}} & \sqrt{\frac{1}{3}} & \frac{1}{2} \sqrt{2-5 s}\\
 \sqrt{\frac{1}{6}-\frac{s}{4}} & \sqrt{\frac{1}{3}} & \frac{\sqrt{2+s}}{2}
 \end{array}
\right),
\end{equation}

\begin{equation}
A4-A5-B9: ~~ |U|  =
\left(
\begin{array}{ccc}
 \sqrt{\frac{2}{3}-s} & \sqrt{\frac{1}{3}} & \sqrt{s} \\
 \sqrt{\frac{2}{15}} & \sqrt{\frac{1}{3}} & 2 \sqrt{\frac{2}{15}} \\
 \sqrt{\frac{1}{5}+s} & \sqrt{\frac{1}{3}} & \sqrt{\frac{7}{15}-s}
\end{array}
\right),
\end{equation}

\begin{equation}
A4-A5-B10: ~~ |U|  =
\left(
\begin{array}{ccc}
 \sqrt{\frac{2}{3}-s} & \sqrt{\frac{1}{3}} & \sqrt{s} \\
 \sqrt{\frac{1}{5}+s} & \sqrt{\frac{1}{3}} & \sqrt{\frac{7}{15}-s}\\
 \sqrt{\frac{2}{15}} & \sqrt{\frac{1}{3}} & 2 \sqrt{\frac{2}{15}}
 \end{array}
\right),
\end{equation}

\begin{equation}
A4-A5-B11: ~~ |U|  =
\left(
\begin{array}{ccc}
 \sqrt{\frac{2}{3}-s} & \sqrt{\frac{1}{3}} & \sqrt{s} \\
 \frac{1}{3} \sqrt{2+12 s} & \sqrt{\frac{1}{3}} & \frac{2}{3} \sqrt{1-3 s} \\
 \frac{1}{3} \sqrt{1-3 s} & \sqrt{\frac{1}{3}} & \frac{1}{3} \sqrt{5+3 s}
\end{array}
\right),
\end{equation}

\begin{equation}
A4-A5-B12: ~~ |U|  =
\left(
\begin{array}{ccc}
 \sqrt{\frac{2}{3}-s} & \sqrt{\frac{1}{3}} & \sqrt{s} \\
 \frac{1}{3} \sqrt{1-3 s} & \sqrt{\frac{1}{3}} & \frac{1}{3} \sqrt{5+3 s}\\
 \frac{1}{3} \sqrt{2+12 s} & \sqrt{\frac{1}{3}} & \frac{2}{3} \sqrt{1-3 s}
 \end{array}
\right),
\end{equation}

\begin{equation}
A3-A4-B2: ~~ |U|  =
\left(
\begin{array}{ccc}
 \sqrt{\frac{2-2s}{3}}  & \sqrt{\frac{1-s}{3}}  & \sqrt{s} \\
 \sqrt{\frac{1+5s}{6}} & \sqrt{\frac{1-s}{3}} & \sqrt{\frac{1-s}{2}} \\
 \sqrt{\frac{1-s}{6}} & \sqrt{\frac{1+2s}{3}} & \sqrt{\frac{1-s}{2}}
\end{array}
\right),
\end{equation}

\begin{equation}
A3-A5-B1: ~~ |U|  =
\left(
\begin{array}{ccc}
 \sqrt{\frac{2-2s}{3}}  & \sqrt{\frac{1-s}{3}}  & \sqrt{s} \\
 \sqrt{\frac{1-s}{6}} & \sqrt{\frac{1+2s}{3}} & \sqrt{\frac{1-s}{2}}\\
 \sqrt{\frac{1+5s}{6}} & \sqrt{\frac{1-s}{3}} & \sqrt{\frac{1-s}{2}}
 \end{array}
\right),
\end{equation}

\begin{equation}
A3-A4-B6: ~~ |U|  =
\left(
\begin{array}{ccc}
 \sqrt{\frac{7-8s}{10}} & \sqrt{\frac{3-2s}{10}} & \sqrt{s} \\
 \sqrt{\frac{2+7s}{10}} & \sqrt{\frac{3-2s}{10}} & \sqrt{\frac{1-s}{2}} \\
 \sqrt{\frac{1+s}{10}} & \sqrt{\frac{2+2s}{5}}  & \sqrt{\frac{1-s}{2}}
\end{array}
\right),
\end{equation}

\begin{equation}
A3-A5-B5: ~~ |U|  =
\left(
\begin{array}{ccc}
 \sqrt{\frac{7-8s}{10}} & \sqrt{\frac{3-2s}{10}} & \sqrt{s} \\
 \sqrt{\frac{1+s}{10}} & \sqrt{\frac{2+2s}{5}}  & \sqrt{\frac{1-s}{2}}\\
 \sqrt{\frac{2+7s}{10}} & \sqrt{\frac{3-2s}{10}} & \sqrt{\frac{1-s}{2}}
 \end{array}
\right).
\end{equation}
Specific mixing angles and Dirac CP phases are listed in Table~\ref{tab:1}~-~Table~\ref{tab:3}.
\begin{table}
\caption{Mixing angles  and Dirac CP phase of  viable leptonic mixing  patterns }
\label{tab:1}       
\begin{tabular}{lllll}
\hline\noalign{\smallskip}
Combinations & $\sin^{2}\theta_{12}$ & $\sin^{2}\theta_{23}$ & $\sin^{2}\theta_{13}$ & ~~$\cos\delta$ \\
\noalign{\smallskip}\hline\noalign{\smallskip}
A1-A4-A7 & $\frac{2+s}{7-7s}$ & $\frac{4-5s}{7-7s}$ & ~~s & $\frac{1-5 s+10 s^2-12 s^3}{2\sqrt{s}\sqrt{10-11 s-8 s^2}\sqrt{12-23 s+10 s^2}}$\\
A1-A5-A6 & $\frac{2+s}{7-7s}$ & $\frac{3-2s}{7-7s}$ & ~~s & $ -\frac{1-5 s+10 s^2-12 s^3}{2\sqrt{s}\sqrt{10-11 s-8 s^2}\sqrt{12-23 s+10 s^2}}$\\
A1-A4-A5 & $\frac{1}{3-3s}$ & ~$\frac{1}{2}$ & ~~s & 0\\
B1-B10-B11 & $\frac{3-11s}{10-10s}$ & ~$\frac{1}{2}$ & ~~s & $\frac{-1-2 s+3 s^2}{4\sqrt{s}\sqrt{21-74s-11s^2}}$\\
B2-B9-B12 & $\frac{3-11s}{10-10s}$ & ~$\frac{1}{2}$ & ~~s & $-\frac{-1-2 s+3 s^2}{4\sqrt{s}\sqrt{21-74s-11s^2}}$\\
B1-B6-B10 & $\frac{13-45s}{45-45s}$ & $\frac{5-9s}{9-9s}$ & ~~s & $\frac{5-31s+90 s^2}{4\sqrt{s}\sqrt{26-90 s}\sqrt{5-9 s}}$\\
B2-B5-B9 & $\frac{13-45s}{45-45s}$ & $\frac{4}{9-9s}$ & ~~s & $-\frac{5-31s+90 s^2}{4\sqrt{s}\sqrt{26-90 s}\sqrt{5-9 s}}$\\
B3-B5-B7 & $\frac{1}{3-3s}$ & $\frac{7}{12-12s}$ & ~~s & $\frac{-1-4 s+12s^2}{\sqrt{7s}\sqrt{5-12s}\sqrt{2-3s}}$\\
A4-B1-B2 & $\frac{1-3s}{3-3s}$ & $\frac{1+2s}{2-2s}$ & ~~s & $\frac{3(1-5 s)\sqrt{s}}{2\sqrt{2}\sqrt{1-3 s}\sqrt{1-2 s-8 s^2}}$\\
A5-B1-B2 & $\frac{1-3s}{3-3s}$ & $\frac{1-4s}{2-2s}$ & ~~s & $-\frac{3(1-5 s)\sqrt{s}}{2\sqrt{2}\sqrt{1-3 s}\sqrt{1-2 s-8 s^2}}$\\
A4-B1-B6 & $\frac{2-5s}{7-7s}$ & $\frac{15+22s}{28-28s}$ & ~~s & $\frac{9+6s+3s^2(-89+10 s)}{2\sqrt{s}\sqrt{(5-2 s)(2-5 s)}\sqrt{(15+22 s)(13-50 s)}}$\\
A5-B2-B5 & $\frac{2-5s}{7-7s}$ & $\frac{13-50s}{28-28s}$ & ~~s & $-\frac{9+6s+3s^2(-89+10 s)}{2\sqrt{s}\sqrt{(5-2 s)(2-5 s)}\sqrt{(15+22 s)(13-50 s)}}$\\
A4-B1-B10 & $\frac{5-25s}{17-17s}$ & $\frac{9+23s}{17-17s}$ & ~~s & $\frac{11+224 s-1365 s^2-150 s^3}{8 \sqrt{10s}\sqrt{9-22 s-115 s^2}\sqrt{3-13 s-10 s^2}}$\\
A5-B2-B9 & $\frac{5-25s}{17-17s}$ & $\frac{8-40s}{17-17s}$ & ~~s & $-\frac{11+224 s-1365 s^2-150 s^3}{8 \sqrt{10s}\sqrt{9-22 s-115 s^2}\sqrt{3-13 s-10 s^2}}$\\
A4-B7-B12 & $\frac{3+4s}{10-10s}$ & $\frac{3-s}{5-5s}$ & ~~s & $-\frac{1+37s-90s^2+24 s^3}{2\sqrt{14s}\sqrt{3-2s-8 s^2}\sqrt{3-7 s+2 s^2}}$\\
A5-B8-B11 & $\frac{3+4s}{10-10s}$ & $\frac{2-4s}{5-5s}$ & ~~s & $\frac{1+37s-90s^2+24 s^3}{2\sqrt{14s}\sqrt{3-2s-8 s^2}\sqrt{3-7 s+2 s^2}}$\\
A4-B6-B10 & $\frac{5}{18-18s}$ & $\frac{5-9s}{9-9s}$ & ~~s & $\frac{7+43s-90s^2}{4\sqrt{s}\sqrt{65-90 s}\sqrt{5-9s}}$\\
A5-B5-B9 & $\frac{5}{18-18s}$ & $\frac{4}{9-9s}$ & ~~s & $-\frac{7+43s-90s^2}{4\sqrt{s}\sqrt{65-90 s}\sqrt{5-9s}}$\\
A4-B9-B11 & $\frac{2-5s}{7-7s}$ & $\frac{4+4s}{7-7s}$ & ~~s & $\frac{1+10s-74s^2+15 s^3}{4\sqrt{s} \sqrt{3-8s-11 s^2} \sqrt{10-29s+10 s^2}}$\\
A5-B10-B12 & $\frac{2-5s}{7-7s}$ & $\frac{3-11s}{7-7s}$ & ~~s & $-\frac{1+10s-74s^2+15 s^3}{4\sqrt{s} \sqrt{3-8s-11 s^2} \sqrt{10-29s+10 s^2}}$\\
A4-B1-B11 & $\frac{7-19s}{23-23s}$ & $\frac{12+20s}{23-23s}$ & ~~s & $\frac{3(5+37 s-317 s^2+19s^3)}{8\sqrt{s}\sqrt{33-74 s-215 s^2} \sqrt{28-83 s+19 s^2}}$\\
A5-B2-B12 & $\frac{7-19s}{23-23s}$ & $\frac{11-43s}{23-23s}$ & ~~s & $-\frac{3(5+37 s-317 s^2+19s^3)}{8\sqrt{s}\sqrt{33-74 s-215 s^2} \sqrt{28-83 s+19 s^2}}$\\
A4-B2-B7 & $\frac{2-5s}{7-7s}$ & $\frac{17+10s}{28-28s}$ & ~~s & $\frac{-1+74s-325s^2+90s^3}{2\sqrt{s}\sqrt{187-536 s-380 s^2}\sqrt{10-29 s+1 s^2}}$\\
A4-B2-B9 & $\frac{9-25s}{29-29s}$ & $\frac{16+20s}{29-29s}$ & ~~s & $\frac{-1+359 s-1735 s^2+225 s^3}{8\sqrt{s}\sqrt{52-131 s-245 s^2}\sqrt{45-134 s+25 s^2}}$\\
A5-B1-B10 & $\frac{9-25s}{29-29s}$ & $\frac{13-49s}{29-29s}$ & ~~s & $-\frac{-1+359 s-1735 s^2+225 s^3}{8\sqrt{s}\sqrt{52-131 s-245 s^2}\sqrt{45-134 s+25 s^2}}$\\

\noalign{\smallskip}\hline
\end{tabular}
\vspace*{15cm}  
\end{table}

\begin{table}
\caption{Mixing angles and Dirac CP phase of  viable leptonic mixing  patterns}
\label{tab:2}       
\begin{tabular}{lllll}
\hline\noalign{\smallskip}
Combinations & $\sin^{2}\theta_{12}$ & $\sin^{2}\theta_{23}$ & $\sin^{2}\theta_{13}$ & ~~$\cos\delta$ \\
\noalign{\smallskip}\hline\noalign{\smallskip}
A4-B10-B11 & ~$\frac{5}{16}$ & ~~$\frac{1}{2}$ & ~~s & $\frac{1+15 s}{2\sqrt{55}\sqrt{s}}$\\
A5-B9-B12 & ~$\frac{5}{16}$ & ~~$\frac{1}{2}$ & ~~s & -$\frac{1+15 s}{2\sqrt{55}\sqrt{s}}$\\
A2-B1-B2 & $\frac{1-3s}{3-3s}$ & ~~$\frac{1}{2}$ & ~~s & ~~~0\\
A2-B1-B10 & $\frac{7-25s}{23-23s}$ & $\frac{11-13s}{23-23s}$ & ~~s & $\frac{16+43 s+10 s^2+75 s^3}{8\sqrt{s} \sqrt{56-193 s-25 s^2}\sqrt{66-133 s+65 s^2}}$\\
A2-B2-B9 & $\frac{7-25s}{23-23s}$ & $\frac{12-10s}{23-23s}$ & ~~s & $-\frac{16+43 s+10 s^2+75 s^3}{8\sqrt{s} \sqrt{56-193 s-25 s^2}\sqrt{66-133 s+65 s^2}}$\\
A2-B1-B11 & $\frac{5-19s}{17-17s}$ & $\frac{8-10s}{17-17s}$ & ~~s & $\frac{12+33 s+10 s^2+57 s^3}{8\sqrt{s}\sqrt{30-109 s-19 s^2} \sqrt{36-73 s+35 s^2}}$\\
A2-B2-B12 & $\frac{5-19s}{17-17s}$ & $\frac{9-7s}{17-17s}$ & ~~s & $-\frac{12+33 s+10 s^2+57 s^3}{8\sqrt{s}\sqrt{30-109 s-19 s^2} \sqrt{36-73 s+35 s^2}}$\\
A3-B1-B10 & $\frac{3-11s}{10-10s}$ & ~~$\frac{1}{2}$ & ~~s & $\frac{(1+3 s)(1-s)}{4\sqrt{s}\sqrt{21-74 s-11 s^2}}$\\
A3-B2-B9 & $\frac{3-11s}{10-10s}$ & ~~$\frac{1}{2}$ & ~~s & $ -\frac{(1+3 s)(1-s)}{4\sqrt{s}\sqrt{21-74 s-11 s^2}}$\\
A7-B1-B2 & $\frac{1-3s}{3-3s}$ & $\frac{7-12s}{12-12s}$ & ~~s & $\frac{1-11s+30 s^2}{2\sqrt{10s}\sqrt{1-3 s}\sqrt{7-12 s}}$\\
A6-B1-B2 & $\frac{1-3s}{3-3s}$ & $\frac{5}{12-12s}$ & ~~s & $-\frac{1-11s+30 s^2}{2\sqrt{10s}\sqrt{1-3 s}\sqrt{7-12 s}}$\\
A7-B1-B6 & $\frac{13-45s}{45-45s}$ & $\frac{5-9s}{9-9s}$ & ~~s & $\frac{5-31 s+90 s^2}{4\sqrt{s}\sqrt{26-90 s}\sqrt{5-9 s}}$\\
A6-B2-B5 & $\frac{13-45s}{45-45s}$ & $\frac{4}{9-9s}$ & ~~s & $-\frac{5-31 s+90 s^2}{4\sqrt{s}\sqrt{26-90 s}\sqrt{5-9 s}}$\\
A7-B1-B11 & $\frac{11-43s}{35-35s}$ & $\frac{4-8s}{7-7s}$ & ~~s & $\frac{9-48 s+217 s^2+78 s^3}{8 \sqrt{2s}\sqrt{33-118 s-43 s^2} \sqrt{3-5 s-2 s^2}}$\\
A6-B2-B12 & $\frac{11-43s}{35-35s}$ & $\frac{3+s}{7-7s}$ & ~~s & $-\frac{9-48 s+217 s^2+78 s^3}{8 \sqrt{2s}\sqrt{33-118 s-43 s^2} \sqrt{3-5 s-2 s^2}}$\\
A7-B2-B9 & $\frac{13-49s}{41-41s}$ & $\frac{24-40s}{41-41s}$ & ~~s & $\frac{25-728 s+1945 s^2-90 s^3}{8\sqrt{2s}\sqrt{91-317 s-98 s^2} \sqrt{51-88 s+5 s^2}}$\\
A6-B1-B10 & $\frac{13-49s}{41-41s}$ & $\frac{17-s}{41-41s}$ & ~~s & $-\frac{25-728 s+1945 s^2-90 s^3}{8\sqrt{2s}\sqrt{91-317 s-98 s^2} \sqrt{51-88 s+5 s^2}}$\\
A7-B3-B5 & $\frac{4+4s}{13-13s}$ & $\frac{8-5s}{13-13s}$ & ~~s & $\frac{-7-73 s+200 s^2-72 s^3}{4\sqrt{s}\sqrt{9-8 s-17 s^2}\sqrt{40-89 s+40 s^2}}$\\
A7-A4-B1 & $\frac{3+5s}{11-11s}$ & $\frac{6-s}{11-11s}$ & ~~s & $\frac{7-131 s+264 s^2-60 s^3}{4 \sqrt{10s}\sqrt{6-13 s+2 s^2}\sqrt{3-s-10s^2}}$\\
A6-A5-B2 & $\frac{3+5s}{11-11s}$ & $\frac{5-10s}{11-11s}$ & ~~s & $-\frac{7-131 s+264 s^2-60 s^3}{4 \sqrt{10s}\sqrt{6-13 s+2 s^2}\sqrt{3-s-10s^2}}$\\
A7-A4-B2 & $\frac{5-s}{17-17s}$ & $\frac{10-19s}{17-17s}$ & ~~s & $\frac{-1+149s-256s^2+36 s^3}{4\sqrt{s}\sqrt{70-113 s-38 s^2} \sqrt{15-23 s+4 s^2}}$\\
A6-A5-B1 & $\frac{5-s}{17-17s}$ & $\frac{7+2s}{17-17s}$ & ~~s & $-\frac{-1+149s-256s^2+36 s^3}{4\sqrt{s}\sqrt{70-113 s-38 s^2} \sqrt{15-23 s+4 s^2}}$\\
A7-A4-B9 & $\frac{4+5s}{14-14s}$ & $\frac{4-2s}{7-7s}$ & ~~s & $\frac{2-70s+149 s^2-45 s^3}{2\sqrt{2s}\sqrt{40-26 s-95 s^2}\sqrt{6-13 s+5 s^2}}$\\
A6-A5-B10 & $\frac{4+5s}{14-14s}$ & $\frac{3-5s}{7-7s}$ & ~~s & $-\frac{2-70s+149 s^2-45 s^3}{2\sqrt{2s}\sqrt{40-26 s-95 s^2}\sqrt{6-13 s+5 s^2}}$\\
A7-A4-B11 & $\frac{4-4s}{14-14s}$ & $\frac{4-8s}{7-7s}$ & ~~s & $\frac{2+50 s-91 s^2+15 s^3}{4\sqrt{s}\sqrt{3-5 s-2 s^2} \sqrt{40-62 s+13 s^2}}$\\
A6-A5-B12 & $\frac{4-4s}{14-14s}$ & $\frac{3+s}{7-7s}$ & ~~s & $-\frac{2+50 s-91 s^2+15 s^3}{4\sqrt{s}\sqrt{3-5 s-2 s^2} \sqrt{40-62 s+13 s^2}}$\\
\noalign{\smallskip}\hline
\end{tabular}
\vspace*{15cm}  
\end{table}

\begin{table}
\caption{Mixing angles and Dirac CP phase of  viable leptonic mixing  patterns}
\label{tab:3}       
\begin{tabular}{lllll}
\hline\noalign{\smallskip}
Combinations & $\sin^{2}\theta_{12}$ & $\sin^{2}\theta_{23}$ & $\sin^{2}\theta_{13}$ & ~~$\cos\delta$ \\
\noalign{\smallskip}\hline\noalign{\smallskip}
A7-A4-B14 & $\frac{5+4s}{16-16s}$ & $\frac{5-4s}{8-8s}$ & ~~s & $\frac{-7-31s+104s^2-48s^3}{2\sqrt{s}\sqrt{55-56s-80s^2} \sqrt{15-32s+16s^2}}$\\
A4-A5-B1 & $\frac{1}{3-3s}$ & $\frac{2+s}{4-4s}$ & ~~s & $\frac{3 \sqrt{s}(-1+2 s)}{\sqrt{2-3 s}\sqrt{4-8s-5s^2}}$\\
A4-A5-B2 & $\frac{1}{3-3s}$ & $\frac{2-5s}{4-4s}$ & ~~s & $-\frac{3 \sqrt{s}(-1+2 s)}{\sqrt{2-3 s}\sqrt{4-8s-5s^2}}$\\
A4-A5-B9 & $\frac{1}{3-3s}$ & $\frac{8}{15-15s}$ & ~~s & $-\frac{1+13s-30 s^2}{4\sqrt{2s}\sqrt{7-15 s}\sqrt{2-3s}}$\\
A4-A5-B10 & $\frac{1}{3-3s}$ & $\frac{7-15s}{15-15s}$ & ~~s & $\frac{1+13s-30 s^2}{4\sqrt{2s}\sqrt{7-15 s}\sqrt{2-3s}}$\\
A4-A5-B11 & $\frac{1}{3-3s}$ & $\frac{4-12s}{9-9s}$ & ~~s & $\frac{1+13 s-30 s^2}{4\sqrt{s}\sqrt{2-3 s}\sqrt{5-12 s-9 s^2}}$\\
A4-A5-B12 & $\frac{1}{3-3s}$ & $\frac{5+3s}{9-9s}$ & ~~s & $-\frac{1+13 s-30 s^2}{4\sqrt{s}\sqrt{2-3 s}\sqrt{5-12 s-9 s^2}}$\\
A3-A4-B2 & ~$\frac{1}{3}$ & ~~$\frac{1}{2}$ & ~~s & $\frac{3 \sqrt{s}}{2\sqrt{2}}$\\
A3-A5-B1 & ~$\frac{1}{3}$ & ~~$\frac{1}{2}$ & ~~s & $-\frac{3 \sqrt{s}}{2\sqrt{2}}$\\
A3-A4-B6 & $\frac{3-2s}{10-10s}$ & ~~$\frac{1}{2}$ & ~~s & $\frac{1+5 s - 6 s^2}{2\sqrt{s}\sqrt{21-38 s+16 s^2}}$\\
A3-A5-B5& $\frac{3-2s}{10-10s}$ & ~~$\frac{1}{2}$ & ~~s & $-\frac{1+5 s - 6 s^2}{2\sqrt{s}\sqrt{21-38 s+16 s^2}}$\\
\noalign{\smallskip}\hline
\end{tabular}
\vspace*{1cm}  
\end{table}

\begin{figure}[tbp]
\centering 
\includegraphics[width=.45\textwidth]{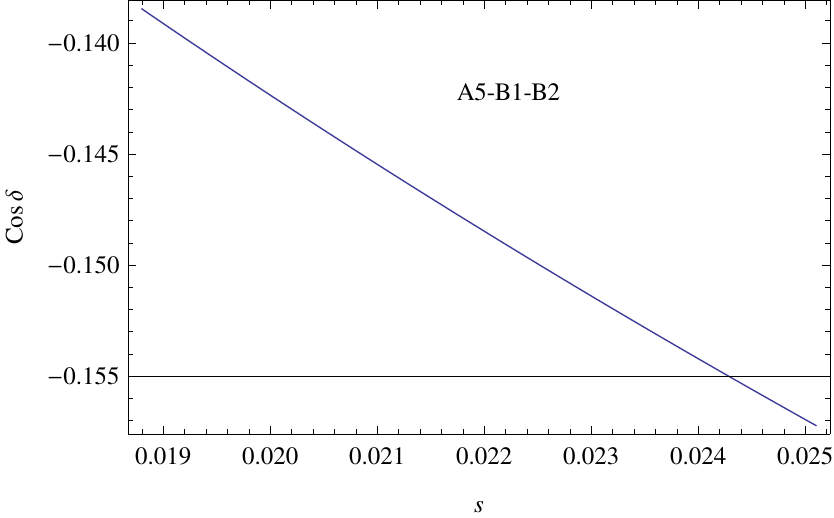}
\hfill
\includegraphics[width=.45\textwidth]{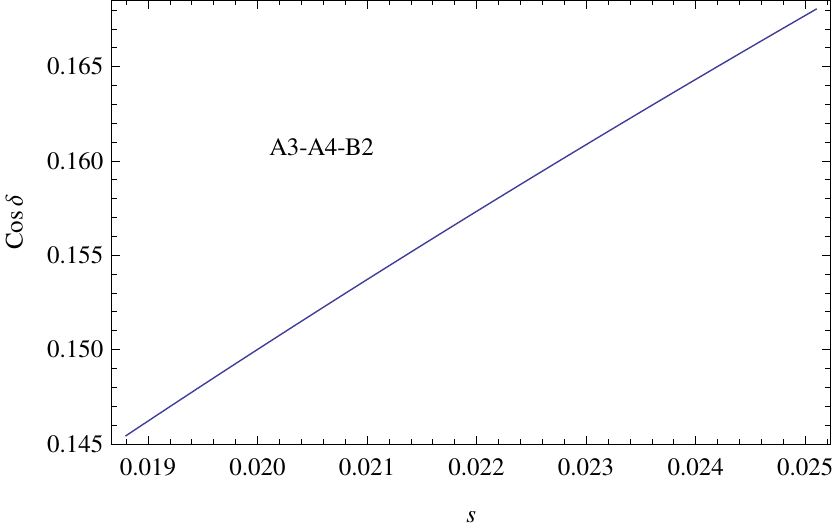}
\caption{\label{fig:1}  $\cos{\delta}$ in viable range of parameter s of perturbative patterns A5-B1-B2 and A3-A4-B2. }
\end{figure}

\begin{figure}[tbp]
\centering 
\includegraphics[width=.45\textwidth]{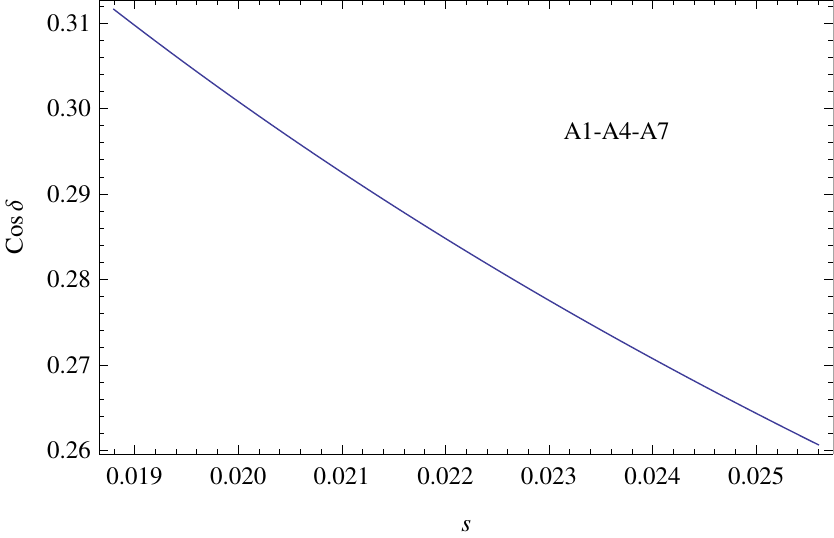}
\hfill
\includegraphics[width=.45\textwidth]{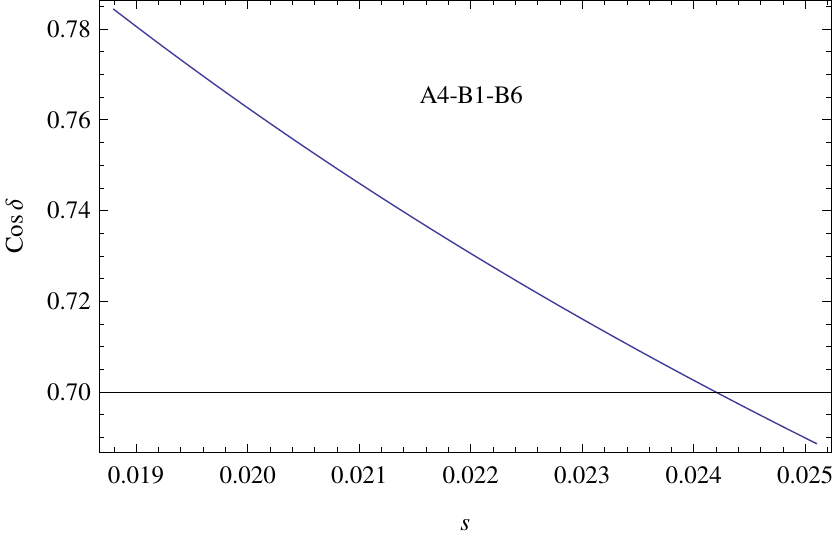}
\caption{\label{fig:2}  $\cos{\delta}$ in viable range of parameter s of nonperturbative patterns A1-A4-A7 and A4-B1-B6. }
\end{figure}

\begin{figure}[tbp]
\centering 
\includegraphics[width=.45\textwidth]{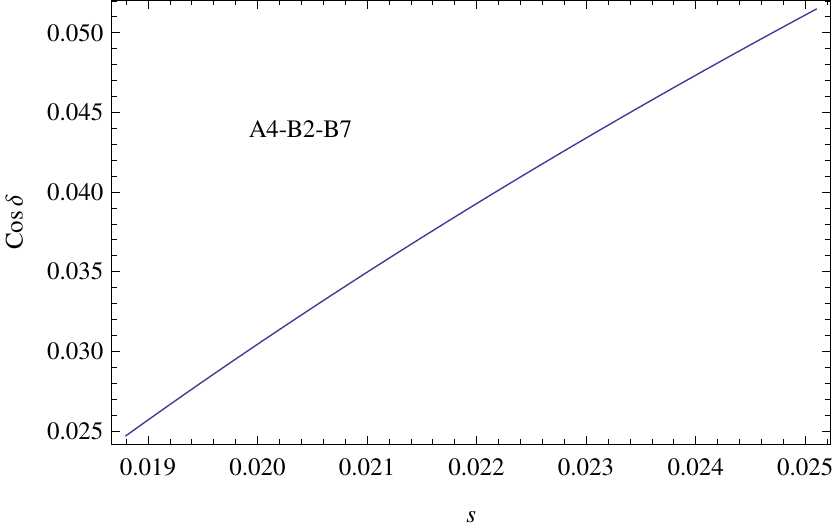}
\caption{\label{fig:3} $\cos{\delta}$ in viable range of parameter s of nonperturbative pattern A4-B2-B7.}
\end{figure}

\begin{figure}[tbp]
\centering 
\includegraphics[width=.45\textwidth]{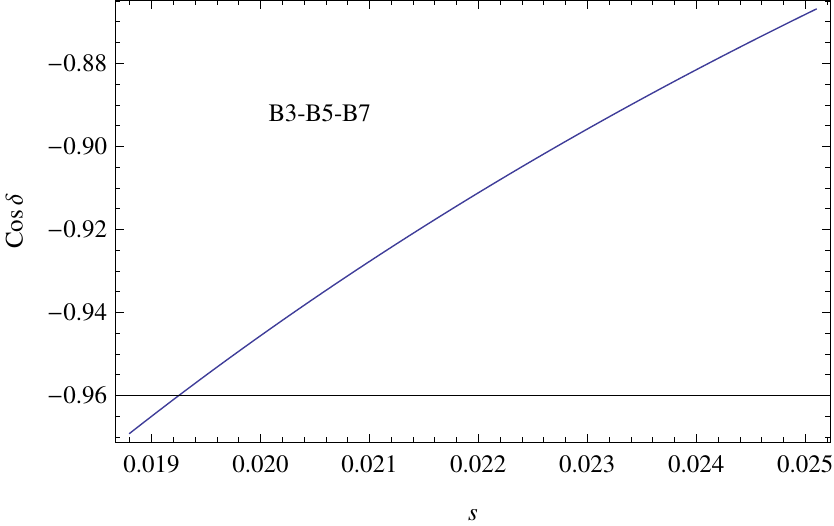}
\hfill
\includegraphics[width=.45\textwidth]{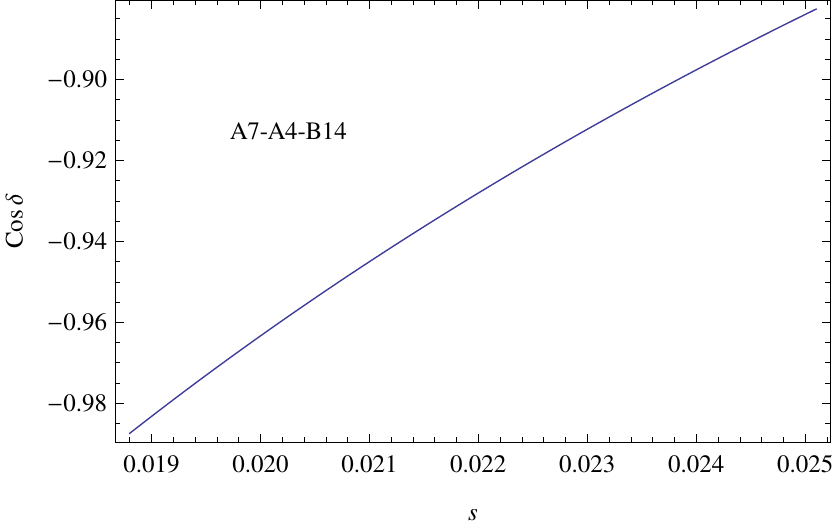}
\caption{\label{fig:4} $\cos{\delta}$ in viable range of parameter s of nonperturbative patterns B3-B5-B7 and A7-A4-B14.}
\end{figure}

\begin{figure}[tbp]
\centering 
\includegraphics[width=.45\textwidth]{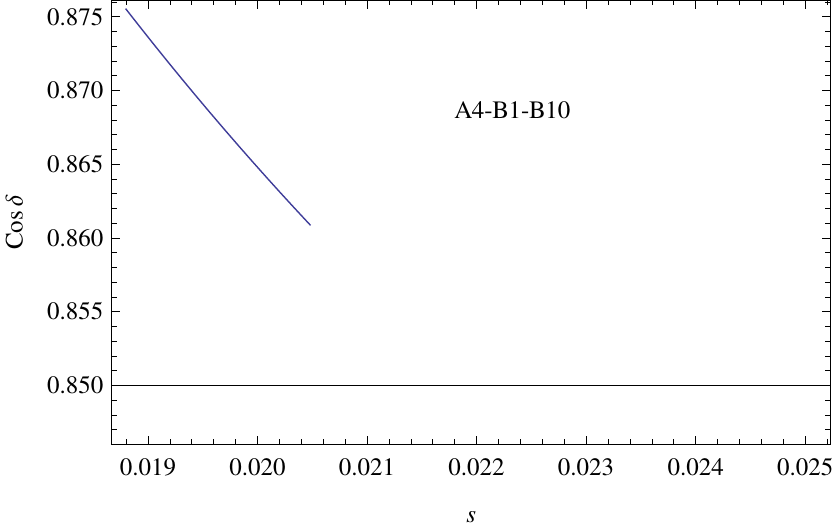}
\hfill
\includegraphics[width=.45\textwidth]{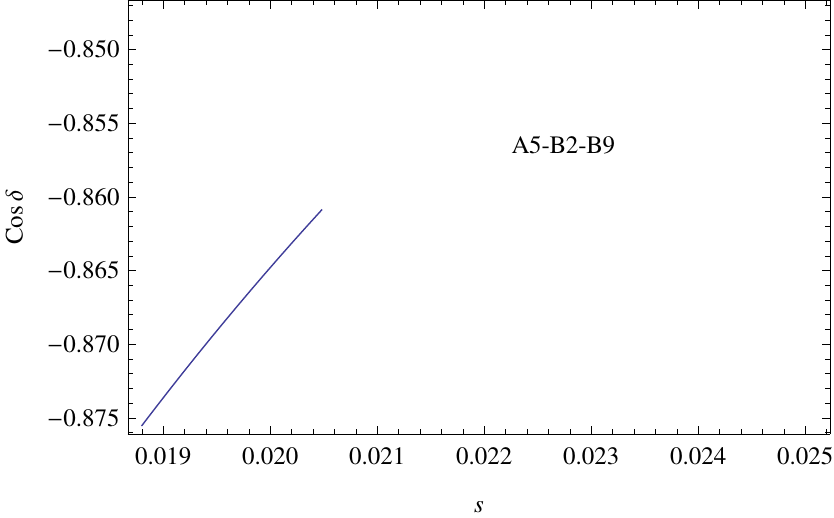}
\caption{\label{fig:5} $\cos{\delta}$ in viable range of parameter s of nonperturbative patterns A4-B1-B10 and A5-B2-B9 .}
\end{figure}

As we can see from the expressions of $|U|$, there are only 6 single combinations, i.e., A1-A4-A5, A2-B1-B2, B3-B5-B7, A4-B2-B7, A7-B3-B5, A7-A4-B14, among which A1-A4-A5 and A2-B1-B2 are $\mu-\tau$ symmetric. The remaining 56 combinations are paired through $\mu-\tau$ interchange. According to the expressions of $\cos\delta$ in tables, all viable patterns can be classified into two groups:  perturbative patterns and nonperturbative patterns. There is a factor $\sqrt{s}$ in $\cos\delta$ of the former pattern which could be obtained from perturbing TBM. So we could call it TBM-like pattern. In contrast, there is a factor $\frac{1}{\sqrt{s}}$ in $\cos\delta$ of the latter pattern which cannot be obtained from perturbing the pattern whose $\theta_{13}$ is zero. There are 8 perturbative patterns, i.e., A1-A4-A5, A4-B1-B2, A5-B1-B2, A2-B1-B2, A4-A5-B1, A4-A5-B2, A3-A4-B2, A3-A5-B1. The remaining 54 patterns are nonperturbative. In more details, as for $\cos\delta$ of perturbative and nonperturbative patterns, we have following
observations:\\
$\bullet$ ~~Because of the factor $\sqrt{s}$, for perturbative patterns, $\cos{\delta}\rightarrow 0$ when $s\rightarrow 0$, see A3-A4-B2, A3-A5-B1
in Table~\ref{tab:3} for example. In contrast, for nonperturbative patterns, because of the factor $\frac{1}{\sqrt{s}}$, $\cos{\delta}\rightarrow \infty$ when $s\rightarrow 0$, see A3-A4-B6 in Table~\ref{tab:3} for example. So for a nonperturbative pattern, small perturbation to $s=0$ would  bring very large variation of Dirac CP phase. We note that this observation has also been obtained in special mixing patterns in Ref.~\cite{43}.\\
$\bullet$  ~~For different perturbative patterns, the difference of $\cos\delta$ is very small. And the variation of $\delta$ is focused on small range around $\pm\frac{\pi}{2}$. We show $\cos\delta$ of A5-B1-B2, A3-A4-B2 as example in Fig.~\ref{fig:1}. In contrast, for two similar nonperturbative patterns, the difference of $\cos\delta$ could be notable. Take A1-A4-A7 and A4-B1-B6 for example, the difference between their  magnitude of elements of mixing matrix is small, while the difference between their $\cos{\delta}$ is large, see Fig.~\ref{fig:2}. \\
$\bullet$  ~~Except the perturbative patterns A2-B1-B2~\eqref{eq:3.31} and A1-A4-A5~\eqref{eq:3.6} as $\mu-\tau$ symmetric combinations, $\cos{\delta}=0$ cannot be obtained at $3\sigma$ level in viable patterns. In nonperturbative patterns, we could obtain $\cos{\delta\simeq 0}$ for A4-B2-B7~\eqref{eq:3.26}, see Fig.~\ref{fig:3}.\\
$\bullet$  ~~Trivial Dirac CP violation phases $\delta=0, \pi$ cannot be obtained at $3\sigma$ level in viable patterns. In nonperturbative patterns, we could obtain $\cos{\delta\simeq -1}$ for B34-B5-B7 and A7-A4-B14, see Fig.~\ref{fig:4}.\\
$\bullet$  ~~The viable range of parameter $ s $ or $ 1/N $ is not reduced obviously for most combinations. The only exceptions are the paired combinations A5-B2-B9 and A4-B1-B10, see Fig.~\ref{fig:5}.

\section{evolutions of mass matrices of special mixing patterns}
In the previous section, we have extracted all viable leptonic mixing  patterns at 3$\sigma$ level from the general combinations of two types of elementary correlations. These viable combinations could be classified into perturbative and nonperturbative patterns. The dependence of $\cos\delta$ on $\sin^{2}\theta_{13}$ of perturbative patterns is different to that of nonperturbative patterns. In this section, we study the dependence of
neutrinos mass matrix on the small mixing parameter $\sin^{2}\theta_{13}$. Our physical motivation to discuss evolutions of mass matrices of special mixing patterns with $\sin^{2}\theta_{13}$ is to examine the stability of a mixing model on the basis of flavor groups and their breaking. We want to know whether a small perturbation of $\sin^{2}\theta_{13}$ would bring about large modifications to the assumption in a flavor model such as alignment of vacuum expectation values of scalar fields. Our discussion is qualitative and is on the basis of special leptonic mixing  patterns. But the conclusions obtained here could be extended to general patterns.

\subsection{Evolutions of mass matrices of perturbative patterns}
The general perturbative  or TBM-like pattern could be expressed as:
\begin{equation}
 |U|  =
\left(
\begin{array}{ccc}
 \sqrt{\frac{2}{3}}  & \sqrt{\frac{1}{3}}  & 0 \\
 \sqrt{\frac{1}{6}} & \sqrt{\frac{1}{3}} & \sqrt{\frac{1}{2}}\\
 \sqrt{\frac{1}{6}} & \sqrt{\frac{1}{3}} & \sqrt{\frac{1}{2}}
 \end{array}
\right)
+\left(
\begin{array}{ccc}
 \epsilon_{11} &  \epsilon_{12}  &  \epsilon_{13} \\
  \epsilon_{21} &  \epsilon_{22} &  \epsilon_{23}\\
  \epsilon_{31} &  \epsilon_{32} &  \epsilon_{33}
 \end{array}
\right),
\end{equation}
where $\epsilon_{ij}$ is a small correction or perturbation of order $\sin\theta_{13}$. And the Dirac CP phase could be written as:
\begin{equation}
\cos\delta=f(\epsilon_{ij}),
\end{equation}
where f is a function dependent regularly on small $\epsilon_{ij}$, which means f(0)=0. Take the perturbative pattern A3-A5-B1 for example, we have:
\begin{equation}
|U|  =
\left(
\begin{array}{ccc}
 \sqrt{\frac{2-2s}{3}}  & \sqrt{\frac{1-s}{3}}  & \sqrt{s} \\
 \sqrt{\frac{1-s}{6}} & \sqrt{\frac{1+2s}{3}} & \sqrt{\frac{1-s}{2}}\\
 \sqrt{\frac{1+5s}{6}} & \sqrt{\frac{1-s}{3}} & \sqrt{\frac{1-s}{2}}
 \end{array}
\right),
\end{equation}
\begin{equation}
\cos{\delta} =-\frac{3 \sqrt{s}}{2\sqrt{2}},
\end{equation}
where $s=\sin^{2}\theta_{13}$.
As for this type of patterns, one can choose TBM pattern as a zeroth-order approximation first, and then take corrections into consideration. These corrections usually come from leptonic interaction or special vacuum expectation of Higgs field or other scalar fields. A large number of papers have constructed leptonic mixing  models for TBM-like patterns following this way, see Refs.~\cite{30,31,32,33,34,35,36} for example.

As an illustrative case, we consider mixing models of pattern A3-A5-B1. The specific mixing matrix of A3-A5-B1 is expressed as:
 \begin{equation}
 U=
 \left(
\begin{array}{ccc}
 \sqrt{\frac{2}{3}}c_{13} & \frac{c_{13}}{\sqrt{3}} & e^{-i\delta } s_{13} \\
 -\frac{1}{\sqrt{6}}-\frac{e^{i\delta}s_{13}}{\sqrt{3}} & \frac{1}{\sqrt{3}}-\frac{e^{i\delta} s_{13}}{\sqrt{6}} & \frac{c_{13}}{\sqrt{2}} \\
 \frac{1}{\sqrt{6}}-\frac{e^{i\delta} s_{13}}{\sqrt{3}} & -\frac{1}{\sqrt{3}}-\frac{e^{i\delta} s_{13}}{\sqrt{6}} & \frac{c_{13}}{\sqrt{2}}
\end{array}
\right)
 \left(
\begin{array}{ccc}
e^{i\alpha_{1}} & 0 & 0 \\
 0 & e^{i\alpha_{2}} & 0 \\
0 & 0 & 1
\end{array}
\right).
 \end{equation}
 In the basis where the mass matrix of charged lepton is diagonal, the mass matrix of Majorana neutrinos is written as:
\begin{equation}
M_{\nu} =Udiag(m_{1},m_{2},m_{3})U^{T}=M^{0}_{\nu}+\delta M_{\nu}.
\end{equation}
The zeroth-order mass matrix of neutrinos $M^{0}_{\nu}$ is of the form:
\begin{align}
M^{0}_{\nu}&=U_{TBM}diag(m_{1},m_{2},m_{3})U^{T}_{TBM} \\ \nonumber
&=\left(
\begin{array}{ccc}
 \frac{2 m_1}{3}+\frac{m_2}{3} & -\frac{m_1}{3}+\frac{m_2}{3} & \frac{m_1}{3}-\frac{m_2}{3} \\
 -\frac{m_1}{3}+\frac{m_2}{3} & \frac{m_1}{6}+\frac{m_2}{3}+\frac{m_3}{2} & -\frac{m_1}{6}-\frac{m_2}{3}+\frac{m_3}{2} \\
 \frac{m_1}{3}-\frac{m_2}{3} & -\frac{m_1}{6}-\frac{m_2}{3}+\frac{m_3}{2} & \frac{m_1}{6}+\frac{m_2}{3}+\frac{m_3}{2}
\end{array}
\right),
\end{align}
where TBM mixing matrix is written as:
\begin{equation}
U_{TBM}=
\left(
\begin{array}{ccc}
\sqrt{ \frac{2}{3}} & \sqrt{ \frac{1}{3}} &0 \\
 -\sqrt{ \frac{1}{6}} & \sqrt{ \frac{1}{3}} & \sqrt{ \frac{1}{2}} \\
\sqrt{ \frac{1}{6}} & -\sqrt{ \frac{1}{3}} & \sqrt{ \frac{1}{2}}
\end{array}
\right).
\end{equation}
Here the Majorana phases have been absorbed into the neutrino mass eigenvalues.
And the correction to zeroth-order mass matrix could be written as:
\begin{equation}
(\delta M_{\nu})_{11}=-\frac{s^{2}_{13}}{3}(2m_{1}+m_{2}-3m_{3}e^{-2i\delta}),
\end{equation}
\begin{equation}
(\delta M_{\nu})_{12}=\frac{1}{3}(1-c_{13})(m_{1}-m_{2})-\frac{\sqrt{2}}{6}s_{13}c_{13}e^{i\delta}(m_{2}+2m_{1})+\frac{\sqrt{2}}{2}s_{13}c_{13}e^{-i\delta}m_{3},
\end{equation}
\begin{equation}
(\delta M_{\nu})_{13}=\frac{1}{3}(1-c_{13})(m_{2}-m_{1})-\frac{\sqrt{2}}{6}s_{13}c_{13}e^{i\delta}(m_{2}+2m_{1})+\frac{\sqrt{2}}{2}s_{13}c_{13}e^{-i\delta}m_{3},
\end{equation}
\begin{equation}
(\delta M_{\nu})_{22}=\frac{s^{2}_{13}}{6}e^{2i\delta}(m_{2}+2m_{1})+\frac{\sqrt{2}}{3}s_{13}e^{i\delta}(m_{1}-m_{2})-\frac{s^{2}_{13}}{2}m_{3},
\end{equation}
\begin{equation}
(\delta M_{\nu})_{23}=\frac{s^{2}_{13}}{6}e^{2i\delta}(m_{2}+2m_{1})-\frac{s^{2}_{13}}{2}m_{3},
\end{equation}
\begin{equation}
(\delta M_{\nu})_{33}=\frac{s^{2}_{13}}{6}e^{2i\delta}(m_{2}+2m_{1})+\frac{\sqrt{2}}{3}s_{13}e^{i\delta}(m_{2}-m_{1})-\frac{s^{2}_{13}}{2}m_{3}.
\end{equation}
In order to construct mixing models, we could choose a special flavor group according to textures of $M^{0}_{\nu}$ and $\delta M_{\nu}$. For example, we choose A$_{4}$ as a flavor group. Following the presentation of A$_{4}$ in the Ref.~\cite{30,37}, we could obtain the general mass matrix of neutrinos~\cite{30}:
\begin{equation}
M_{\nu}=m_{0}
\left(
\begin{array}{ccc}
 a+\frac{2}{3}b_{1} & c-\frac{1}{3}b_{3} & d-\frac{1}{3}b_{2} \\
  c-\frac{1}{3}b_{3} & d+\frac{1}{3}b_{2} & a-\frac{2}{3}b_{1}\\
 d-\frac{1}{3}b_{2}  & a-\frac{2}{3}b_{1} &  c+\frac{1}{3}b_{3}
\end{array}
\right),
\end{equation}
where $m_{0}$, a, b$_{i}$, c, d come from the vacuum expectation values of scalar fields as singlet or triplet of A$_{4}$.
Then employing the following equation:
\begin{equation}
m_{0}
\left(
\begin{array}{ccc}
 a+\frac{2}{3}b_{1} & c-\frac{1}{3}b_{3} & d-\frac{1}{3}b_{2} \\
  c-\frac{1}{3}b_{3} & d+\frac{1}{3}b_{2} & a-\frac{2}{3}b_{1}\\
 d-\frac{1}{3}b_{2}  & a-\frac{2}{3}b_{1} &  c+\frac{1}{3}b_{3}
\end{array}
\right)
=M^{0}_{\nu}+\delta M_{\nu},
\end{equation}
we could obtain the special alignment of vacuum expectation values of scalar fields. Now one may consider that we have obtained a neutrinos mixing model that could  interpret the special mixing pattern. However, we should answer the important question what the physical origin of special alignment of vacuum expectations is. At the level of zeroth-order mass matrix, the alignment of vacuum expectations could be obtained by residual symmetries such as $Z_{2}$, $C_{3}$ after spontaneous breaking of A$_{4}$, see Ref.~\cite{39} for example. However, we should answer the further question whether the assumption of such mixing model is stable under the perturbation of small mixing parameters. Let us examine the evolution of neutrinos mass matrix with $s_{13}$.
At present, absolute mass scales of neutrinos and ordering of neutrino masses are unknown. The effects of $s_{13}$ on corrections to neutrinos mass matrix include following cases:\\
Case~1~~$|m_{1}|\sim|m_{2}|\sim|m_{3}|$.
In this case, corrections of order $s^{2}_{13}|m_{i}|$ could be omitted, the correction matrix could be expressed as:
\begin{equation}
\delta M_{\nu}\backsimeq
\left(
\begin{array}{ccc}
0& \epsilon_{2} &\epsilon_{2} \\
 \epsilon_{2} & \epsilon_{1} & 0\\
\epsilon_{2} & 0 & -\epsilon_{1}
\end{array}
\right),
\end{equation}
where $\epsilon_{i}$ is of the form:
\begin{equation}
\label{eq:4.9}
\begin{array}{c}
\epsilon_{1}=\frac{\sqrt{2}}{3}s_{13}e^{i\delta}(m_{1}-m_{2}),\\
\epsilon_{2}=\frac{\sqrt{2}}{6}s_{13}c_{13}e^{i\delta}(3e^{-2i\delta}m_{3}-m_{2}-2m_{1}).\\
\end{array}
\end{equation}
Case~2~~ $|m_{1}|<|m_{2}|<|m_{3}|$. In this case, corrections of order $ s^{2}_{13}|m_{3}|$ cannot be omitted.
The correction matrix could be expressed as:
\begin{equation}
\delta M_{\nu}\backsimeq
\left(
\begin{array}{ccc}
\epsilon'_{1}& \epsilon'_{2} &\epsilon'_{2} \\
 \epsilon'_{2} & \epsilon'_{3} &\epsilon'_{5}\\
\epsilon'_{2} & \epsilon'_{5} & \epsilon'_{4}
\end{array}
\right),
\end{equation}
where $\epsilon'_{i}$ is of the form:
\begin{equation}
\label{eq:4.9}
\begin{array}{c}
\epsilon'_{1}=m_{3}s^{2}_{13}e^{-2i\delta},~~\epsilon'_{5}=-\frac{s^{2}_{13}}{2}m_{3},\\
\epsilon'_{3}=\frac{\sqrt{2}}{3}s_{13}e^{i\delta}(m_{1}-m_{2})-\frac{s^{2}_{13}}{2}m_{3},\\
\epsilon'_{4}=\frac{\sqrt{2}}{3}s_{13}e^{i\delta}(m_{2}-m_{1})-\frac{s^{2}_{13}}{2}m_{3},
\end{array}
\end{equation}
and the expression of $\epsilon'_{2}$ is the same as $\epsilon_{2}$ in Case 1.\\
Case~3~~ $|m_{3}|<|m_{1}|<|m_{2}|$. In this case, corrections of order $s^{2}_{13}|m_{i}|$ could also be omitted, the correction matrix is the same as that in Case 1.\\
From corrections of elements of the mass matrix, we can see that small $s_{13}$ corresponds to small modification of the structure of the neutrinos mass matrix in the Case 1 and Case 3. However, small $s_{13}$ would change the neutrinos mass matrix notably in the Case 2. Especially, for example, if $|m_{3}|\sim 10|m_{i}|$ (i=1, 2), the correction $\epsilon'_{2}$ brought by small $s_{13}$ is large compared with the zeroth-order element.
Therefore, only when neutrinos mass scales of $|m_{i}|$(i=1,2,3) approximate or when neutrinos masses are in inverted ordering, the corrections are small and could be treated as perturbation. So the structure of neutrinos the mass matrix of perturbative pattern is dependent on the mass scales of neutrinos. A small variation of $s_{13}$ may correspond to large modifications to alignment of vacuum expectation values in mixing models of Case 2. Therefore, in general, the robustness of a mixing model on the basis of flavor group is dependent on neutrinos mass scales. Similar observations have been obtained in Refs.~\cite{38,46} with different perturbative methods.

\subsection{Evolutions of mass matrices of nonperturbative patterns}

Different to the perturbative patterns, there is no unified formula to depict the nonperturbative mixing patterns. Because the dependence of $\cos\delta$ on $\sin^{2}\theta_{13}$ is not continuous when $\theta_{13}\rightarrow0$, we cannot obtain a
 nonperturbative pattern from perturbing TBM pattern. However, as a theoretical program, we could still choose a zeroth-order mixing pattern whose $\theta_{13}\neq 0$, and then take corrections into consideration, see Ref.~\cite{29} for example. On the other hand, we could survey various flavor groups and choose special residual symmetries to construct viable mixing models without extra corrections. Following this program, orders of viable discrete flavor groups are very large. And viable mixing patterns on the basis of large finite groups are usually TM$_{2}$, see Refs.~\cite{40,41,42}.

 For both of these programs, in order to study the stability of the flavor model of a special mixing pattern,  variation of lepton mass matrix should be examined when the perturbation of small mixing parameter is considered. As an illustrative example, we discuss the evolution of neutrinos mass matrix of pattern A5-B9-B12. The mixing matrix of A5-B9-B12 is written as:
 \begin{equation}
\left(
\begin{array}{ccc}
 \frac{\sqrt{11}}{4}c_{13} & \frac{\sqrt{5}}{4}c_{13} & e^{-i\delta}s_{13} \\
 -\sqrt{\frac{5}{32}}-\sqrt{\frac{11}{32}} e^{i\delta}s_{13} & \sqrt{\frac{11}{32}}-\sqrt{\frac{5}{32}} e^{i\delta}s_{13} & \frac{c_{13}}{\sqrt{2}}\\
  \sqrt{\frac{5}{32}}-\sqrt{\frac{11}{32}} e^{i\delta}s_{13} &-\sqrt{\frac{11}{32}}-\sqrt{\frac{5}{32}} e^{i\delta}s_{13} & \frac{c_{13}}{\sqrt{2}}
\end{array}
\right),
 \end{equation}
 where~$\cos\delta$~is listed in Table ~\ref{tab:2}.
 In the basis where the mass matrix of charged lepton is diagonal, elements of mass matrix of Majorana neutrinos are written as:
\begin{equation}
(M_{\nu})_{11}=\frac{11}{16}c^{2}_{13}m_{1}+\frac{5}{16}c^{2}_{13}m_{2}+s^{2}_{13}e^{-2i\delta}m_{3},
\end{equation}
\begin{equation}
(M_{\nu})_{12}=-\frac{\sqrt{2}}{32}(\sqrt{55}c_{13}+11s_{13}c_{13}e^{i\delta})m_{1}+\frac{\sqrt{2}}{32}(\sqrt{55}c_{13}-11s_{13}c_{13}e^{i\delta})m_{2}+\frac{\sqrt{2}}{2}s_{13}c_{13}e^{-i\delta}m_{3},
\end{equation}
\begin{equation}
(M_{\nu})_{13}=\frac{\sqrt{2}}{32}(\sqrt{55}c_{13}-11s_{13}c_{13}e^{i\delta})m_{1}-\frac{\sqrt{2}}{32}(\sqrt{55}c_{13}+11s_{13}c_{13}e^{i\delta})m_{2}+\frac{\sqrt{2}}{2}s_{13}c_{13}e^{-i\delta}m_{3},
\end{equation}
\begin{equation}
(M_{\nu})_{22}=\frac{1}{32}(\sqrt{5}+\sqrt{11}s_{13}e^{i\delta})^{2}m_{1}+\frac{1}{32}(\sqrt{5}-\sqrt{11}s_{13}e^{i\delta})^{2}m_{2}+\frac{1}{2}c^{2}_{13}m_{3},
\end{equation}
\begin{equation}
(M_{\nu})_{23}=\frac{1}{32}(-5+11s^{2}_{13}e^{2i\delta})m_{1}+\frac{1}{32}(-11+5s^{2}_{13}e^{2i\delta})m_{2}+\frac{1}{2}c^{2}_{13}m_{3},
\end{equation}
\begin{equation}
(M_{\nu})_{33}=\frac{1}{32}(\sqrt{5}-\sqrt{11}s_{13}e^{i\delta})^{2}m_{1}+\frac{1}{32}(\sqrt{5}+\sqrt{11}s_{13}e^{i\delta})^{2}m_{2}+\frac{1}{2}c^{2}_{13}m_{3}.
\end{equation}
As the Ref~\cite{46}, we introduce following parameters to describe the structure of mass matrix, i.e.,
\begin{equation}
\varepsilon_{a}=\frac{(M_{\nu})_{12}-(M_{\nu})_{13}}{(M_{\nu})_{12}+(M_{\nu})_{13}}=
\frac{\sqrt{110}(m_{2}-m_{1})}{-11\sqrt{2}s_{13}e^{i\delta}(m_{1}+m_{2})+16\sqrt{2}s_{13}e^{-i\delta}m_{3}},
\end{equation}

\begin{equation}
\varepsilon_{b}=\frac{(M_{\nu})_{22}-(M_{\nu})_{33}}{(M_{\nu})_{22}+(M_{\nu})_{33}}=
\frac{2\sqrt{55}s_{13}e^{i\delta}(m_{1}-m_{2})}{(5+11s^{2}_{13}e^{2i\delta})(m_{1}+m_{2})+16c^{2}_{13}m_{3}}.
\end{equation}
Through the derivative
\begin{equation}
\frac{d\varepsilon_{a}}{ds_{13}}\simeq\frac{C}{s^{2}_{13}}=
\frac{-\sqrt{110}(m_{2}-m_{1})}{[-11\sqrt{2}e^{i\delta}(m_{1}+m_{2})+16\sqrt{2}e^{-i\delta}m_{3}]s^{2}_{13}},
\end{equation}
\begin{equation}
\frac{d\varepsilon_{b}}{ds_{13}}\simeq
\frac{2\sqrt{55}e^{i\delta}(m_{1}-m_{2})}{(5+11s^{2}_{13}e^{2i\delta})(m_{1}+m_{2})+16c^{2}_{13}m_{3}},
\end{equation}
we can see that the structure parameter $\varepsilon_{a}$ dependents on small parameter $s^{2}_{13}$ singularly when $s^{2}_{13}\rightarrow 0$.
Thus, except the case where the coefficient $|C|\ll1$, a small perturbation of mixing parameter $s_{13}$ would bring about a notable variation of structure parameter $\varepsilon_{a}$. So the structure of mass matrix is stable only in special mass ordering of neutrinos such as normal hierarchy, i.e., $|m_{1}|\ll|m_{2}|\ll|m_{3}|$ or quasidegeneracy, i.e., $|m_{1}|\simeq|m_{2}|<|m_{3}|$.
Therefore, in general, the robustness of the structure of neutrinos mass matrix of a mixing model of nonperturbative pattern is also dependent on the mass scales of neutrinos. Accordingly, a small variation of $s_{13}$ may correspond to large modifications to alignment of vacuum expectation values in a mixing model on the basis of flavor groups in the special mass ordering of neutrinos.

\section{Summary}
We have obtained all viable patterns of leptonic mixing  matrix on the basis of combinations of elementary correlations of elements of mixing matrix. The elementary correlations include type-A: $|U_{\alpha i}|=|U_{\beta j}|$; type-B: $|U_{\alpha i}|=2|U_{\beta j}|$. There are 9 viable type-A correlation  and 14 viable type-B correlation at $3\sigma$ level of mixing parameters. With the help of $\mu-\tau$ interchange of mixing matrix, we examined all viable combinations of elementary correlations. We obtained 83 viable  combinations of two elementary correlations and 62 viable combinations of three elementary correlations. A combination of three elementary correlations could determine a neutrinos mixing pattern.
All these viable patterns could be classified into two groups: perturbative patterns and nonperturbative patterns. The former can be obtained by perturbing TBM. The latter cannot be obtained from any $\theta_{13}=0$ pattern. The Dirac CP violating phases of perturbative patterns are compared with those of nonperturbative patterns through function $\cos\delta(s)$ and figures.

In order to study the robustness of leptonic mixing  models on the basis of flavor groups, evolutions of mass matrix of neutrinos have been discussed via special perturbative and nonperturbative mixing patterns. We found that the mass matrix of neutrinos is stable under the small variation of mixing parameter $\sin\theta_{13}$ only in special mass ordering of neutrinos. In general cases, a small variation of $\sin\theta_{13}$ may correspond to large modifications to alignment of vacuum expectation values in a mixing model on the basis of flavor groups. Therefore, small but nonzero $\sin\theta_{13}$ brings more stringent constraint on leptonic mixing  models on the basis of flavor groups than usual views. A successful mixing model should not only predict viable mixing parameters but also be stable under the small variation of mixing parameter.

Finally, we note that a recent paper~\cite{44} has discussed viable patterns on the basis of type-A correlation without discussing evolutions of neutrinos mass matrices. Our work incudes both type-A and type-B correlation. We find that, in 62 viable patterns, only 3 are just from  combinations of correlations of type-A. So general methods and results of our work are different from theirs.

\acknowledgments
This work is supported by the National Natural Science Foundation of China under the Grant No. 11405101.

\end{document}